\documentclass[12pt,preprint]{aastex}
\usepackage{multirow}
\usepackage{graphicx}        % For eps figures, newer & more powerfull
\usepackage{natbib}         % For citations: redefine \cite commands
\usepackage{amssymb}        % useful mathematical symbols
\usepackage[fleqn]{amsmath}
\usepackage{color}           % For color text: \color command
\usepackage{url}             % For breaking URLs easily trough lines
\usepackage{epstopdf}
\graphicspath{ {FIG/} }
\usepackage{graphicx}
\usepackage{amsmath}
\usepackage{float}

\shorttitle{}
\shortauthors{}

\newcommand{\be}{{\bf e}}

\newcommand{\bg}{{\bf g}}
\newcommand{\bu}{{\bf u}}
\newcommand{\Ri}{{\rm Ri}}
\newcommand{\Ree}{{\rm Re}}
\newcommand{\Pe}{{\rm Pe}}

\newcommand{\bnabla}{{\mathbf{\nabla}}}
\usepackage{mathtools}

\begin{document}


\title{Turbulent transport by diffusive stratified shear flows: from local to global models. III. A closure model. }


\author{Logithan KULENTHIRARAJAH$^{1,2}$ \& Pascale GARAUD$^3$}
\affil{$^1$Institut de Recherche en Astrophysique et Plan\'etologie (IRAP), Toulouse, France  \\
	$^2$Universit\'e F́\'ed́\'erale Toulouse Midi-Pyr\'en\'ees (UFTMiP)\\
$^3$Department of Applied Mathematics and Statistics, Baskin School of Engineering, University of California Santa Cruz, 1156 High Street, Santa Cruz CA 95060}


\begin{abstract}
Being able to account for the missing mixing in stellar radiative zones is a key step toward a better understanding of stellar evolution.  Zahn (1974) argued that thermally diffusive shear-induced turbulence might be responsible for some of this mixing. In Part I and Part II of this series of papers we showed that Zahn's (1974, 1992) mixing model applies when the properties of the turbulence are local. But we also discovered limitations of the model when this locality condition fails, in particular near the edge of a turbulent region. In this paper, we propose a second-order closure model for the transport of momentum and chemical species by shear-induced turbulence in strongly stratified, thermally diffusive environments (the so-called low P\'eclet number limit), which builds upon the work of Garaud \& Ogilvie (2005). Comparison against direct numerical simulations (DNSs) shows that the model is able to predict the vertical profiles of the mean flow and of the stress tensor (including the momentum transport) in diffusive shear flows, often with a reasonably good precision, and at least within a factor of order unity in the worst case scenario. The model is sufficiently simple to be implemented in stellar evolution codes, and all the model constants have been calibrated against DNSs. While significant limitations to its use remain (e.g. it can only be used in the low P\'eclet number, slowly rotating limit), we argue that it is more reliable than most of the astrophysical prescriptions that are used in stellar evolution models today. 
\end{abstract}
	
	
	

	%% in the electronic edition to the object data ilable at the data centers  

\section{Introduction}
\label{sec:intro}

First discussed in the context of stellar astrophysics by \citet{Zahn1974} \citep[see also][]{EndalSofia78,Zahn92,Maeder95,MaederMeynet2000}, diffusive shear instabilities -- also called secular shear instabilities -- are now considered by most stellar evolution codes \citep{Pinsonneaultal1989,Hegeral2000,Eggenberger2008,Paxton13} and can be a significant source of mixing in some phases of stellar evolution. This is by contrast with standard shear instabilities (in which perturbations are assumed to be adiabatic), which cannot usually develop in stellar radiative zones owing to the overwhelmingly stabilizing effect of the thermal stratification. Indeed, the well-known energy-based criterion for instability in the adiabatic case is \citep{Richardson1920}  
\begin{equation}
J  = \frac{N^2}{S^2} < J_c  , 
\label{eq:Rich}
\end{equation}
where $N$ is the local Brunt-V\"ais\"al\"a frequency, $S$ is the local shearing rate, and $J_c$ is a constant of order unity. It is rarely satisfied in radiative zones since $J$ is usually much larger than one. \citet{Zahn1974} argued that thermal diffusion can however mitigate the stabilizing impact of thermal stratification, to the extent that the relevant criterion for instability should no longer be given by (\ref{eq:Rich}) but instead by 
\begin{equation}
J {\rm Pr}  < (J{\rm Pr})_c  ,
\label{eq:Zahncrit}
\end{equation}
where ${\rm Pr} = \nu/\kappa_T$ is the Prandtl number ($\nu$ is the kinematic viscosity and $\kappa_T$ is the thermal diffusivity) and where $(J{\rm Pr})_c $ is a number of order $10^{-3}$. Since ${\rm Pr} \sim 10^{-9} - 10^{-6}$ in stellar interiors, Zahn's criterion would suggest that diffusive shear instabilities are possible for $J$ up to $10^3 - 10^6$, which is more in line with the range of observed values in stars.

\citet{Zahn92} later put forward a simple model for turbulent transport by such instabilities, arguing that the typical scale $l$ of the turbulent eddies is the largest possible scale for which $J {\rm Pe}_l$ is of order unity, where 
\begin{equation}
{\rm Pe}_l = \frac{S l^2}{\kappa_T} 
\end{equation}
is the eddy-scale P\'eclet number (i.e. the ratio of the advective timescale to the thermal diffusion timescale). That scale will hereafter be referred to as the Zahn scale, and is given by  
\begin{equation}
J \frac{S l_{\rm Z}^2}{\kappa_T}  = (J{\rm Pe})_c \Rightarrow l_{\rm Z} = \sqrt{ \frac{ (J {\rm Pe})_c \kappa_T }{JS} }  ,
\end{equation}
where $(J {\rm Pe})_c$ is a constant of order one. An estimate for the turbulent diffusion coefficient can then readily be obtained by simple dimensional analysis \citep{Zahn92}:  
\begin{equation}
D_{\rm turb} = \beta S l_{\rm Z}^2 = \beta  (J {\rm Pe})_c \frac{ \kappa_T }{J} \equiv C \frac{ \kappa_T }{J}  ,
\label{eq:ZahnDturb}
\end{equation}
where $\beta$ and $C$ are thus both constants as well. Applying Zahn's model for diffusive shear instabilities therefore involves using (\ref{eq:ZahnDturb}) whenever (\ref{eq:Zahncrit}) is satisfied. It is important to note, however, that this model is purely {\it local}, and is not expected to apply when the Zahn scale becomes commensurate with the shear lengthscale, or with any other macroscopic system scale, such as the distance to the surface or to the nearest convective zone for instance. 

In the past few years, tremendous progress has been made using both theory and Direct Numerical Simulations (DNSs) to test Zahn's model, and to gain a more detailed understanding of the properties of diffusive shear instabilities. DNSs being limited to modeling a very small region of a star, they most commonly use the Boussinesq approximation for gases \citep{SpiegelVeronis1960}, in which one assumes the fluid is in hydrostatic and thermal equilibrium, and where perturbations evolve according to  
\begin{eqnarray}
	&&\frac{\partial\bu}{\partial t} + \bu \cdot \bnabla \bu  = -\frac{\bnabla p}{\rho_0}  + \alpha g T \be_z + \nu \nabla^2\bu \label{eq:momentum}, \\
	&&\frac{\partial T}{\partial t}+\bu\cdot\bnabla T + u_z (T_{0z}-T_{{\rm ad},z}) =  \kappa_T\nabla^2T\label{eq:heat},  \\
	&&\bnabla\cdot\bu  =  0\label{eq:continuity},
	\end{eqnarray}
where $\bu = (u_x,u_y,u_z)$ is the fluid velocity, $p$ is the pressure perturbation away from hydrostatic equilibrium, $T$ is the temperature perturbation away from radiative equilibrium (which has a temperature gradient $T_{0z}$, and an adiabatic temperature gradient $T_{{\rm ad},z}$), $\rho_0$ is the mean background density of the fluid in this region, $\bg = - g \be_z$ is the gravity (and $\be_z$ is the local vertical direction), and $\alpha$ is the thermal expansion coefficient. The quantities $T_{0z}$, $T_{{\rm ad},z}$, $\rho_0$, $g$, $\alpha$, $\nu$ and $\kappa_T$ are assumed to be constant. 

 Since the instability requires thermal diffusion to proceed, one of the major theoretical advances in this field has been the formal derivation of the so-called the Low P\'eclet Number (LPN) equations by \citet{Lignieres1999}, who showed that whenever the local eddy-scale P\'eclet number is small, the evolution of the temperature perturbations is well-approximated by solving 
 \begin{equation}
 u_z (T_{0z}-T_{{\rm ad},z}) =  \kappa_T\nabla^2T\label{eq:LPN1}
  \end{equation}
for $T$ instead of evolving (\ref{eq:heat}). In this limit, the temperature perturbations and the vertical velocity are therefore slaved to one-another, and $T$ can be solved for knowing $u_z$. This has a number of important consequences. For instance, it can easily be shown by taking a horizontal average of (\ref{eq:LPN1}) that temperature perturbations must have zero mean if mass is conserved, which implies that the temperature stratification within the star remains unaffected by LPN turbulence. It is also obvious from (\ref{eq:LPN1}) that the system dynamics now only ever depend on the ratio  $(T_{0z}-T_{{\rm ad},z})/\kappa_T$, which formally shows how a large thermal diffusivity can damp the effect of stratification. Finally, the total number of independent unknowns in the LPN limit is reduced by one compared with the full equations, which both saves computing time in DNSs, and permits the derivation of interesting theoretical results, among which is the formal identification of criterion (\ref{eq:Zahncrit}) as an energy stability criterion \citep{Bischoff13,Garaudal15}.

Using DNSs within the LPN limit, \citet{PratLignieres13}, \citet{PratLignieres14}, \citet{Garaudal15}, \citet{Pratal2016}), \citet{GaraudKulen16}, \citet{Garaudal17} (Paper I hereafter) and Gagnier \& Garaud (2018) (Paper II hereafter) tested various aspects of Zahn's model. As reviewed in Papers I and II, the general conclusion of these studies is that Zahn's model performs remarkably well as long as its fundamental assumption, namely that the turbulence is local, applies. In particular, the works of Prat and collaborators agree with those of Garaud and collaborators in estimating the critical threshold for instability in (\ref{eq:Zahncrit}) to be 
\begin{equation}
(J {\rm Pr})_c \simeq 0.007 .
\end{equation}
Paper I also found that the turbulent mixing model (\ref{eq:ZahnDturb}) fares relatively well whenever the local assumption is verified, with a coefficient 
\begin{equation}
C \simeq 0.08
\end{equation}
 for the turbulent diffusivity of a passive scalar and similarly for the turbulent viscosity. On the other hand, both Paper I and Paper II demonstrated  
the need for an extension of Zahn's model in the limit where the local assumption is no longer valid (i.e. when the Zahn scale becomes commensurate with any system scale). Paper II also revealed that a region which does not satisfy the instability criterion (\ref{eq:Zahncrit}) may still undergo substantial turbulent mixing if it is adjacent to one which does -- a shear-driven analog to convective overshoot. Finally, Paper II  demonstrated that in the strongly stratified limit, diffusive shear instabilities are intrinsically nonlinear and therefore exhibit hysteretic properties. As a result, a region that satisfies (\ref{eq:Zahncrit}) is not necessarily turbulent; whether it is or not depends on whether it is linearly unstable, and if not, then it depends on the evolutionary history of the shear layer.

	In order to go beyond a strictly local description of turbulent transport in stratified shear flows, we propose in this paper a second-order turbulence closure model. This type of approach is well established, and first consists in deriving exact evolution equations for the second-order moments of the equations (which include the Reynolds stress tensor and the turbulent buoyancy fluxes for instance). These equations in turn depend on third-order correlations, for which a closure model is then proposed. The closure model considered here is based on the one first put forward by \citet{Ogilvie2003} and applied to unstratified shear-induced turbulence by \citet{GaraudOgilvie2005} and to turbulent convection by \citet{Garaudal2010}. We now extend it to study stratified shear flows in the LPN limit. The construction of the model closely follows the steps outlined in \citet{Garaudal2010}, then proceeds to apply the LPN asymptotic expansion of \citet{Lignieres1999} to the closure equations. This is done in Section \ref{sec:closure}. Section \ref{sec:local}  explores the local properties of the model, and constrains its closure parameters against the existing DNSs of stratified plane Couette flows presented in Paper I. Section \ref{sec:Kolmo} compares the model predictions with the DNSs of body-forced stratified Kolmogorov flow presented in \citet{Garaudal15} and \citet{GaraudKulen16}. Section \ref{sec:tanh} compares them to the DNSs of Paper II, and addresses model inadequacies through the introduction of an additional turbulent diffusivity term in the Reynolds stress equations. In Section \ref{sec:L}, we propose a simple model for the turbulent eddy scale, which is a necessary ingredient of the closure model. We conclude in Section \ref{sec:ccl} by summarizing the findings of this paper, and by discussing caveats and possible future extensions of the model. 
	
	\section{A second-order turbulence closure model for diffusive shear instabilities}
        \label{sec:closure}

			\subsection{Reynolds stress evolution equations}
			\label{sec:stresseq}
			
			 We start from the momentum, thermal energy and mass conservation equations (\ref{eq:momentum}) to (\ref{eq:continuity}), now written for simplicity using Einstein's convention of summation over repeated indices, and 
			 generically non-dimensionalized assuming a given velocity scale $U_0$, lengthscale $L_0$ and temperature scale $T_0 = (T_{0z} - T_{{\rm ad},z}) L_0$. We have:
			\begin{eqnarray}
			&&\partial_t u_i +\partial_k(u_iu_k) = -\partial_ip +{\rm Ri} T\delta_{iz} + \frac{1}{\rm Re}\partial^2_{kk} u_i  \left( + \hat F(z) \delta_{ix} \right) \label{eq:cm1},\\
			&&\partial_t T+ \partial_k(u_k T) + u_z = \frac{1}{\rm Pe}\partial^2_{kk}T  \label{eq:cmt1},\\
			&&\partial_k u_k = 0\label{eq:cmc1},
			\end{eqnarray}
			where the indices $i$ and $k$ can be $x$, $y$ or $z$ (and similarly for the index $j$ later on), and where 
			\begin{eqnarray}
			&& \Ri= \frac{N^2 L^2_0  }{U_0^2}, \mbox{  where  } N^2 = \alpha g  (T_{0z} - T_{{\rm ad},z}),  \label{eq:Ri} \\ 
	&& \Ree=\frac{U_0L_0}{\nu}, \\
	&& \Pe=\frac{U_0L_0}{\kappa_T} \label{eq:Pe}.
	\end{eqnarray}
	                  The quantities $\Ree$, $\Ri$ and $\Pe$ constructed in this manner can be thought of respectively as the Richardson number, Reynolds number and P\'eclet number for the flow. Since we will be comparing the closure model to different sets of DNSs that each use different normalizations, we leave the actual selection of $U_0$ and $L_0$ unspecified for now. Note that we have allowed for the existence of a non-dimensional body force to drive the shear flow in (\ref{eq:cm1}), but the latter is not required for wall-bounded shear flows (e.g. Paper I) or homogeneous shear flows in the shearing sheet approximation \citep{Pratal2016}. 
						
			 As in \citet{Ogilvie2003} we now consider a suitable averaging procedure, which could be an average over realizations, or over a small lengthscale and/or short timescales, and which must commute with all differential operators ($\partial_i$ and $\partial_t$). This average is denoted with an overbar. Applying it to equations (\ref{eq:cm1}) to (\ref{eq:cmc1}) yields the mean equations:
			\begin{eqnarray}
			&&\partial_t \bar u_i + \partial_k(\overline{u_iu_k})=-\partial_i\bar {p} +\Ri\bar {T}\delta_{iz} + \frac{1}{\Ree}\partial^2_{kk} \bar u_i \left(  + \hat F(z)\delta_{ix} \right) \label{eq:cm2},\\
			&&\partial_t\bar {T} + \partial_k(\overline{u_k T}) + \bar {u}_z = \frac{1}{\Pe}\partial^{2}_{kk}\bar {T}\label{eq:cmt2},\\
			&&\partial_k\bar u_k = 0\label{eq:cmc2}.
			\end{eqnarray}
			We then separate all variables into mean and fluctuating parts such as $f=\bar {f} + f' $
			where by construction $\overline{f'} =0$. Using this decomposition in  (\ref{eq:cm1}) to (\ref{eq:cmc1}) and subtracting the mean equations  (\ref{eq:cm2}) to (\ref{eq:cmc2}) we obtain the evolution equations for the fluctuations:
			\begin{eqnarray}
			&&\partial_t u_i' + \bar u_k\partial_ku_i' + u_k' \partial_k \bar u_i + \partial_k( u_i' u_k ') -\partial_k(\overline{u_iu_k})=-\partial_ip' +\Ri T'\delta_{iz} + \frac{1}{\Ree}\partial^2_{kk}u_i'\label{eq:cm3},\\
			&&\partial_tT' + \bar u_k \partial_k T' + u'_k \partial_k \bar T + u'_z + \partial_k (u'_k T')  - \partial_k(\overline{u_kT}) = \frac{1}{\Pe}\partial^{2}_{kk}T'\label{eq:cmt3},\\
			&&\partial_k u_k'= 0\label{eq:cmc3}.
			\end{eqnarray}
			From these we can derive the exact equations for the evolution of the averages of quadratic quantities such as $R_{ij}= u_i'u_j'$ (the Reynolds stress), $F_i=u_i' T'$  (the turbulent temperature flux) and $Q = T'^2$ as follow:
			\begin{eqnarray}
			&& \partial_t\bar R_{ij} + \bar u_k\partial_k\bar R_{ij} +  \bar R_{jk}\partial_k\bar u_i+\bar R_{ik}\partial_k\bar u_j - \Ri \bar F_i\delta_{jz} - \Ri \bar F_j\delta_{iz} - \frac{1}{\Ree}\partial^2_{kk} \bar R_{ij} \nonumber  \\ &&= -\overline{u_j'\partial_ip'} -\overline{u_i'\partial_jp'} - \overline{u_k'\partial_kR_{ij}} - \frac{2}{\Ree}   \overline{\partial_k u_i' \partial_{k} u_j'}  \label{eq:cm4}, \\
			&& \partial_t\bar F_{i} + \bar u_k\partial_k\bar F_{i} + \bar F_{k}\partial_k\bar u_i + \bar R_{ik}\partial_k\bar {T} - \Ri\bar {Q}\delta_{iz} + \bar R_{iz} - \frac{1}{2}\left(\frac{1}{\Ree}+\frac{1}{\Pe}\right)\partial^2_{kk}\bar F_i \nonumber  \\ && = -\overline{T'\partial_ip'} - \overline{T' \partial_k R_{ik} + u_i'\partial_kF_k} + \frac{1}{2}\left(\frac{1}{\Ree}-\frac{1}{\Pe}\right)\left(\partial_k(\overline{T'\partial_k u_i' - u_i'\partial_kT')}\right) \nonumber \\ && - \left(\frac{1}{\Ree}+\frac{1}{\Pe}\right) (\overline{\partial_k u_i'\partial_kT'}) \label{eq:cmt4} ,			\\ 
			&&			\partial_t\bar {Q} + \bar u_k \partial_k \bar Q + 2\bar F_k\partial_k \bar {T} + 2\bar F_z - \frac{1}{\Pe}\partial^2_{kk}\overline{Q} =  -\overline{u_k'\partial_k Q} - \frac{2}{\Pe}\overline{(\partial_k T')^2} \label{eq:Tm4}.
			\end{eqnarray}
			These equations are identical to those of \citet{Garaudal2010} aside from the contribution associated with the assumed constant background temperature gradient which manifests itself in the terms $\bar R_{iz}$ in equation (\ref{eq:cmt4}) and $2\bar F_z$ in equation (\ref{eq:Tm4}). The left-hand-side contains quantities that can be expressed exactly in terms of $\bar R_{ij}$, $ \bar F_{i}$, $\bar u_i$, $\bar {T}$ and $\bar {Q}$ or their gradients. On the right-hand-side are correlations involving the pressure fluctuations, triple correlations between fluctuating quantities and dissipative terms involving microscopic diffusion. These terms cannot directly be expressed as functions of  $\bar R_{ij}$, $ \bar F_{i}$ and $\bar Q$ and must be modeled as part of the closure problem. 
			
			\subsection{Proposed second-order closure model}
		
			We adopt a simple closure model of \citet{Garaudal2010} \citep[see also][]{Ogilvie2003,GaraudOgilvie2005}, applied to the set of equations (\ref{eq:cm4}) to (\ref{eq:Tm4}):  
%			\begin{eqnarray}
%			&&\begin{split}
%			\partial_t\overline{R_{kn}} + & \overline{R_{nj}}\partial_j\overline{u_k}+\overline{R_{kj}}\partial_j\overline{u_n} - \Ri_F\overline{F_k}\delta_{nz} - \Ri_F\overline{F_n}\delta_{kz}\\ &=-\frac{C_1R^{1/2}}{L} \overline{R_{kn}} -\frac{C_2R^{1/2}}{L}\left( \overline{R_{kn}}-\frac{1}{3}R\delta_{kn}\right) - C_{pk}\Ri_FF_k - C_{pn}\Ri_FF_n  \\&-\frac{C_{\nu}\overline{R_{kn}}}{\Ree_FL^2}+ \frac{1}{\Ree_F}\partial_{zz}\overline{R_{kn}}\label{eq:cm5},
%			\end{split}	
%			\\
%			&&\begin{split}
%			\partial_t\overline{F_{k}} + &\overline{F_{j}}\partial_j\overline{u_k} + \overline{R_{kj}}\partial_j\overline{T} - \Ri_F\overline{Q}\delta_{kz} + \overline{R_{kn}}\delta_{nz} - \frac{1}{2}\left(\frac{1}{\Ree_F}+\frac{1}{\Pe_F}\right)\partial^2_{jj}\overline{F_k}\\&=-\frac{C_6R^{1/2}}{L} \overline{F_{k}} -\frac{1}{2L^2}\left(\frac{1}{\Ree_F}+\frac{1}{\Pe_F}\right)C_{\nu k}\overline{F_k}\label{eq:cmt5},
%			\end{split}
%			\\ 
%			&&\begin{split}
%			\partial_t\overline{Q} + 2\overline{F_j}\partial_j \overline{T} + 2\overline{F_z} - \frac{1}{\Pe_F}\overline{\partial^{2}_{jj}T'} =  -2\overline{u_j'\partial_j Q} + \frac{2}{\Pe_F}\overline{(\partial_j T')^2} \label{eq:Tm5},
%			\end{split}
%			\end{eqnarray}
			\begin{eqnarray}
			&&\begin{split}
			\partial_t\bar R_{ij} + &  \bar u_k\partial_k\bar R_{ij}  +  \bar R_{ik}\partial_k\bar u_j+\bar R_{jk}\partial_k\bar u_i - \Ri \bar F_i\delta_{jz} - \Ri \bar F_j\delta_{iz} - \frac{1}{\Ree}\partial_{kk}\bar R_{ij} \\ &=-\frac{C_1\bar R^{1/2}}{L} \bar R_{ij} -\frac{C_2\bar R^{1/2}}{L}\left( \bar R_{ij}-\frac{1}{3}\bar {R}\delta_{ij}\right)-\frac{C_{\nu}\bar R_{ij}}{\Ree L^2}\label{eq:cm5},
			\end{split}	
			\\
			&&\begin{split}
			\partial_t\bar F_{i} + & \bar u_k\partial_k\bar F_{i} + \bar F_{k}\partial_k\bar u_i + \bar R_{ik}\partial_k\bar {T} - \Ri \bar {Q}\delta_{iz} + \bar R_{iz} - \frac{1}{2}\left(\frac{1}{\Ree}+\frac{1}{\Pe}\right)\partial^2_{kk}\bar F_i \\&=-\frac{C_6\bar R^{1/2}}{L} \bar F_{i} -\frac{1}{2L^2}\left(\frac{1}{\Ree}+\frac{1}{\Pe}\right)C_{\nu \kappa}\bar F_i\label{eq:cmt5},
			\end{split}
			\\ 
			&&\begin{split}
			\partial_t\bar {Q} + \bar u_k \partial_k \bar Q + 2\bar F_k\partial_k \bar {T} + 2\bar F_z - \frac{1}{\Pe} \partial^{2}_{kk} \overline{Q}  = - \frac{C_7\bar {R}^{1/2}}{L} \bar {Q} - \frac{C_\kappa}{\Pe L^2}\bar {Q} \label{eq:Tm5},
			\end{split}
			\end{eqnarray}
			where $C_1$, $C_2$, $C_6$, $C_7$, $C_{\nu}$, $C_k$ and $C_{\nu k}$ are assumed to be positive dimensionless universal constants, where $\bar R$ is the trace of the Reynolds stress tensor (which is also an estimate of the square of the turbulent rms velocity), 
			\begin{equation}
			\bar R = \bar R_{xx}  + \bar R_{yy} + \bar R_{zz} \end{equation}
			 and $L$ is a characteristic lengthscale of the dominant turbulent eddies. The terms containing $C_1$ represent the decay of the turbulent kinetic energy through the turbulent cascade on the eddy turnover timescale of $L/\bar {R}^{1/2}$. The terms containing $C_2$ redistribute the turbulent energy through the different component of $\bar R_{ij}$, and model the expected isotropization of the turbulence on a similar timescale. The terms with $C_6$ and $C_7$ represent the turbulent decay of temperature fluctuations. $C_\nu$ is introduced to model the viscous dissipation when an efficient turbulent cascade does not form at low Reynolds number, in which case the dissipation rate is directly proportional to the viscosity. In a similar way, we introduce the last terms of equation (\ref{eq:cmt4}) and (\ref{eq:Tm4}) containing $C_\kappa$ and $C_{\nu \kappa}$. 
			
			\subsection{Closure model in LPN limit}
			
			The equations derived so far could in principle be used to model shear-induced turbulence in high P\'eclet number stratified shear flows, but we do not recommend it. Indeed, stratified turbulence in that limit can be strongly influenced by gravity waves, whose long-range effects cannot be modeled within the context of this semi-local closure approach. However, gravity waves are not supported in the LPN limit \citep{Lignieres1999}, providing us with hope that the closure model can be successfully applied here.
			
			In order to derive a formal asymptotic limit of the closure model equations at low P\'eclet numbers, we apply the procedure of \cite{Lignieres1999}, assuming a decomposition of the form $ \bar R_{ij}=\bar R_{ij}^{(0)} + \Pe \bar R_{ij}^{(1)} + O(\Pe^2)$, $ \bar F_{i}=\bar F_{i}^{(0)} + \Pe \bar F_{i}^{(1)} + O(\Pe^2)$ and $ \bar Q=\bar {Q}^{(0)} + \Pe \bar {Q}^{(1)} + O(\Pe^2)$.  To the lowest order in $\Pe$ we have:
			\begin{eqnarray}
			\partial^2_{kk} \bar F_x^{(0)} = \frac{C_{\nu \kappa}}{L^2}\bar F_x^{(0)},
			\qquad
			\partial^2_{kk} \bar F_z^{(0)} = \frac{C_{\nu \kappa}}{L^2}\bar F_z^{(0)},
			\qquad
			\partial^2_{kk} \bar Q^{(0)} = \frac{C_\kappa}{L^2}\bar {Q}^{(0)},
			\end{eqnarray}
			which do not have any bounded or periodic solutions. As a result for our particular model setup, $\bar F_x^{(0)}=\bar F_z^{(0)}=\bar {Q}^{(0)}=0$. To the next order in $\Pe$ we have:
			\begin{eqnarray}
			\partial^2_{kk} \bar F_x^{(1)} = 2\bar R_{xz}^{(0)}+\frac{C_{\nu \kappa}}{L^2}\bar F_x^{(1)},
			\qquad
			\partial^2_{kk} \bar F_z^{(1)} =2\bar R_{zz}^{(0)}-\Ri\Pe\bar Q^{(1)}+ \frac{C_{\nu \kappa}}{L^2}\bar F_z^{(1)},
			\qquad
			\partial^2_{kk} \bar Q^{(1)} = \frac{C_\kappa}{L^2}\bar Q^{(1)},
			\end{eqnarray}
			which implies that $\bar Q^{(1)}=0$ for the same reasons. Combining the remaining $\bar F_x^{(1)}$ and $ \bar F_z^{(1)}$ equations with (\ref{eq:cm5}) - (\ref{eq:Tm5}) yields:
			\begin{eqnarray}
			&&\begin{split}
			\partial_t\bar R_{ij} + &\bar u_k\partial_k\bar R_{ij} +  \bar R_{ik}\partial_k\bar u_j+\bar R_{jk}\partial_k\bar u_i - \Ri\Pe (\bar f_i\delta_{jz} + \bar f_j\delta_{iz} ) - \frac{1}{\Ree}\partial_{kk}\bar R_{ij}\\&=-\frac{C_1\bar R^{1/2}}{L} \bar R_{ij} -\frac{C_2\bar R^{1/2}}{L}\left( \bar R_{ij}-\frac{1}{3}\bar {R}\delta_{ij}\right) -\frac{C_{\nu}\bar R_{ij}}{\Ree L^2}\label{eq:cm6},
			\end{split}	
			\\
			&&\begin{split}
			\bar R_{iz} - \frac{1}{2}\partial^2_{kk}\bar f_i=-\frac{1}{2L^2}C_{\nu \kappa}\bar f_i\label{eq:cmt6},
			\end{split}
			\end{eqnarray}
			where we have defined $f_i=F_i/\Pe$ for simplicity. We recover the fundamental property of LPN flows, namely that their dynamics only depend on the product of the Richardson and P\'eclet numbers $\Ri\Pe$.

			In all that follows, we now take the average to imply a horizontal average. With this, we have $\bar u_z = 0$ by mass conservation, and all the horizontal derivatives disappear so
						\begin{eqnarray}
			&&\begin{split}
			\partial_t\bar R_{xx} +  2\bar R_{xz}\partial_z\bar {u}_x - \frac{1}{\Ree}\partial^2_{zz}\bar R_{xx} =-(C_1+C_2)\frac{\bar {R}^{1/2}}{L} \bar R_{xx}  + \frac{C_2\bar {R}^{3/2}}{3L}  -\frac{C_{\nu}\bar R_{xx}}{\Ree L^2}\label{eq:dcm1},
			\end{split}	
			\\
			&&\begin{split}
			\partial_t\bar R_{yy} - \frac{1}{\Ree }\partial^2_{zz}\bar R_{yy}=-(C_1+C_2)\frac{\bar {R}^{1/2}}{L} \bar R_{yy}  + \frac{C_2\bar {R}^{3/2}}{3L}  -\frac{C_{\nu}\bar R_{yy}}{\Ree L^2}\label{eq:dcm2},
			\end{split}	
			\\
			&&\begin{split}
			\partial_t\bar R_{zz} -  2\Ri\Pe\bar f_z  - \frac{1}{\Ree}\partial^2_{zz}\bar R_{zz} =-(C_1+C_2)\frac{\bar {R}^{1/2}}{L} \bar R_{zz}  + \frac{C_2\bar {R}^{3/2}}{3L} -\frac{C_{\nu}\bar R_{zz}}{\Ree L^2}\label{eq:dc3},
			\end{split}	
			\\
			&&\begin{split}
			\partial_t\bar R_{xz} +  \bar R_{zz}\partial_z\bar {u}_x- \Ri \Pe \bar f_x- \frac{1}{\Ree}\partial^2_{zz}\bar R_{xz} =-(C_1+C_2)\frac{\bar {R}^{1/2}}{L} \bar R_{xz}  -\frac{C_{\nu}\bar R_{xz}}{\Ree L^2}\label{eq:dcm4},
			\end{split}	
			\\
			&&\begin{split}
			\partial_t\bar R_{yz} -  \Ri \Pe \bar f_y  - \frac{1}{\Ree}\partial^2_{zz}\bar R_{yz}=- (C_1+C_2) \frac{\bar R^{1/2}}{L} \bar R_{yz}  -\frac{C_{\nu}\bar R_{yz}}{\Ree L^2}\label{eq:dcm5},
			\end{split}	
			\\
			&&\begin{split}
			\partial_t\bar R_{xy} +  \bar R_{xz}\partial_z\bar {u}_x - \frac{1}{\Ree }\partial^2_{zz}\bar R_{xy}=-(C_1+C_2) \frac{\bar {R}^{1/2}}{L} \bar R_{xy}  -\frac{C_{\nu}\bar R_{xy}}{\Ree L^2}\label{eq:dcm6},
			\end{split}	
			\\
			&&\begin{split}
			\bar R_{xz} - \frac{1}{2}\partial^2_{zz}\bar f_x=-\frac{1}{2L^2}C_{\nu \kappa}\bar f_x\label{eq:dcmt1},
			\end{split}
			\\
			 &&\begin{split}
			\bar R_{yz} - \frac{1}{2}\partial^2_{zz}\bar f_y=-\frac{1}{2L^2}C_{\nu \kappa}\bar f_y\label{eq:dcmt2},
			\end{split}
			\\
			&&\begin{split}
			\bar R_{zz} - \frac{1}{2}\partial^2_{zz}\bar f_z=-\frac{1}{2L^2}C_{\nu \kappa}\bar f_z\label{eq:dcmt3}.
			\end{split}
			\end{eqnarray}
			Note how the temperature correlations $\bar Q$ and the original model constants $C_6$, $C_7$ and $C_\kappa$ drop out of the LPN closure model. Also note that the $\bar f_y$ and $\bar R_{yz}$ equations decouple from the others. Since they are not forced, these quantities must eventually always decay away. In what follows, we now assume them to be zero. In addition, the quantity $\bar R_{xy}$ is not involved in any equation aside from its own evolution equation. We can therefore ignore it until needed for some particular purpose. The remaining coupled system of equations reduces to equations (\ref{eq:dcm1})-(\ref{eq:dcm4}) as well as (\ref{eq:dcmt1}) and  (\ref{eq:dcmt3}), together with the mean flow equation (\ref{eq:cm2}) where $\bar T = 0$ in the LPN limit (see Section \ref{sec:intro}).

			\section{Local properties of the closure model, and parameter fitting}
			\label{sec:local}
		
			\subsection{Local properties of the model}		
			\label{sec:local1}

			It is informative to investigate what the closure model predictions are for homogeneous stratified shear flows. To do so, we assume that the background shear $\partial_z \bar u_x \equiv S$ is known and constant, and neglect all other derivatives in the closure equations accordingly. In this section, we ignore the body force, and select the following non-dimensionalization: $S^{-1}$ as unit time, $L_z$ as unit length (where $L_z$ is the height of the computational domain considered), and $L_z (T_{0z} - T^{\rm ad}_z)$ as unit temperature. This redefines the Richardson, P\'eclet and Reynolds numbers to be $\Ri \equiv J = N^2/S^2$, $\Pe \equiv {\rm Pe}_S = SL_z^2 / \kappa_T$ and $\Ree \equiv {\rm Re}_S = S L_z^2 / \nu$. The closure equations then reduce to 
			\begin{eqnarray}
			&&	 2 \bar R_{xz} s = - \frac{C_1+C_2}{L} \bar R^{1/2}  \bar R_{xx}  + \frac{C_2}{3L} \bar R^{3/2} -  \frac{1}{{\rm Re}_S} \frac{C_\nu}{L^2} \bar R_{xx}, \\ 
			&&	 0  = - \frac{C_1+C_2}{L} \bar R^{1/2}  \bar R_{yy}  + \frac{C_2}{3L} \bar R^{3/2} -  \frac{1}{{\rm Re}_S} \frac{C_\nu}{L^2} \bar R_{yy}, \\ 
			&&	 - 2 J {\rm Pe}_S  \bar f_z = - \frac{C_1+C_2}{L} \bar R^{1/2}  \bar R_{zz}  + \frac{C_2}{3L} \bar R^{3/2} -  \frac{1}{{\rm Re}_S} \frac{C_\nu}{L^2} \bar R_{zz}, \\ 
			&& \bar R_{zz}  s - J {\rm Pe}_S \bar f_x =- \frac{C_1+C_2}{L} \bar R^{1/2}  \bar R_{xz}  -  \frac{1}{{\rm Re}_S} \frac{C_\nu}{L^2} \bar R_{xz}, \\ 
			&& \bar R_{xz} =  - \frac{C_{\nu\kappa}}{2 L^2} \bar f_x, \\
			&& \bar R_{zz} =  - \frac{C_{\nu\kappa}}{2 L^2} \bar f_z ,
			\end{eqnarray}
			where $L$ is now implicitly given in units of $L_z$, and $s = {\rm sign}(S)$. The two flux equations can straightforwardly be solved for $\bar f_x$ and  $\bar f_z$, and substituted into the Reynolds stress equations. Furthermore, we can construct an evolution equation for $\bar R$ by summing the $\bar R_{xx}$, $\bar R_{yy}$ and $\bar R_{zz}$ equations. Once this is done, both $\bar R_{xx}$ and $\bar R_{yy}$ decouple from the system, which can be reduced to three coupled equations only (for $\bar R$, $\bar R_{zz}$, and $\bar R_{xz}$): 
			\begin{eqnarray}
			%&&	 2 \bar R_{xz} s = - \frac{C_1+C_2}{L} \bar R^{1/2}  \bar R_{xx}  + \frac{C_2}{3L} \bar R^{3/2} -  \frac{1}{{\rm Re}_S} \frac{C_\nu}{L^2} \bar R_{xx} \label{eq:localeqs1}  \\ 
			&&	 4 J{\rm Pe}_S \frac{L^2}{C_{\nu\kappa}}  \bar R_{zz} = - \frac{C_1+C_2}{L} \bar R^{1/2}  \bar R_{zz}  + \frac{C_2}{3L} \bar R^{3/2} -  \frac{1}{{\rm Re}_S} \frac{C_\nu}{L^2} \bar R_{zz} ,\label{eq:localeqszz}  \\ 
			&&	 2 \bar R_{xz} s + 4 J {\rm Pe}_S \frac{L^2}{C_{\nu\kappa}} \bar R_{zz}= - \frac{C_1}{L} \bar R^{3/2}    -  \frac{1}{{\rm Re}_S} \frac{C_\nu}{L^2} \bar R ,\label{eq:localeqsR}  \\ 
			&& \bar R_{zz}  s + 2J {\rm Pe}_S \frac{L^2}{C_{\nu\kappa}}  \bar R_{xz} =- \frac{C_1+C_2}{L} \bar R^{1/2}  \bar R_{xz}  -  \frac{1}{{\rm Re}_S} \frac{C_\nu}{L^2} \bar R_{xz} .
			\label{eq:localeqsxz} 
			\end{eqnarray}
			% Checked 
			We see that two parameters naturally emerge, namely $\sigma = J {\rm Pe}_S L^2 / C_{\nu\kappa}$, which controls the stratification, and $\epsilon = C_\nu / {\rm Re}_S L^2$, which controls macroscopic viscous damping on the scale $L$. Furthermore, if we define $\lambda = \bar R^{1/2}/L$, $r_{zz} = \bar R_{zz}/\bar R$ and similarly for $r_{xz}$, we get the simpler-looking equations
			\begin{eqnarray}
			&&	 4 \sigma  r_{zz} = - (C_1+C_2) \lambda r_{zz}  + \frac{C_2}{3} \lambda - \epsilon r_{zz}, \label{eq:simpleeqszz}  \\ 
			&&	 2 r_{xz} s + 4 \sigma r_{zz}= - C_1 \lambda  -  \epsilon,  \label{eq:simpleeqsR}  \\ 
			&&  r_{zz}  s + 2\sigma r_{xz} =- (C_1+C_2) \lambda r_{xz}  -  \epsilon r_{xz} ,
			\label{eq:simpleeqsxz} 
			\end{eqnarray}
		        which can be combined into a single cubic equation for $\lambda$: 
			\begin{equation} 
			\frac{2C_2}{3} \lambda  = \left\{  4\sigma \left[  \left(C_1 + \frac{C_2}{3} \right) \lambda  + \epsilon \right] + (C_1 \lambda  + \epsilon) \left( (C_1 + C_2)\lambda  + \epsilon\right) \right\} \left[ 2\sigma + (C_1+C_2)\lambda  + \epsilon \right] .
			%Checked 
			\end{equation}
		        For general values of the parameters $\sigma$ and $\epsilon$, this equation must be solved numerically, and from there, the values of $\bar R$ and of all the components of the stress tensor can be recovered. In some limits, analytical solutions also exist. 
			
\begin{figure}[!ht]
				\begin{center}
					\includegraphics[width=\textwidth]{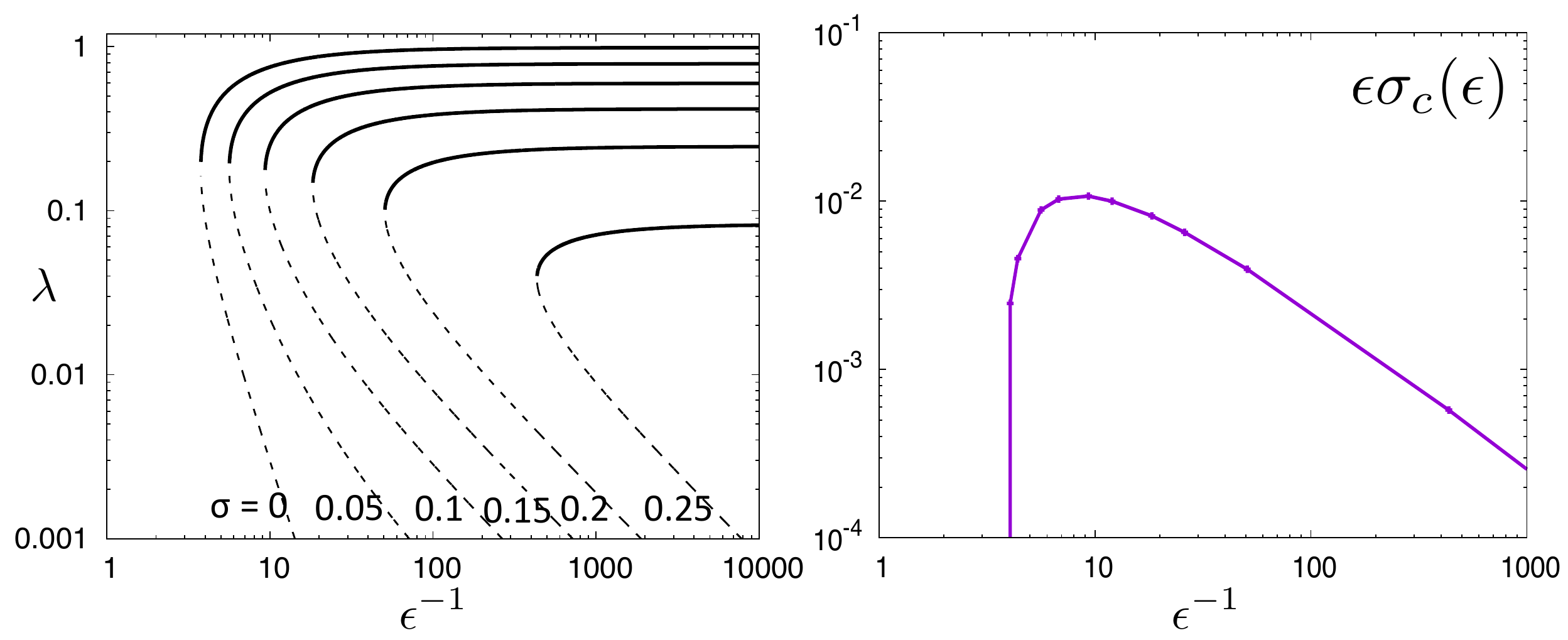}
					\caption{Left: Numerical solutions for $\lambda_1$ (repelling turbulent branch, dashed line) and $\lambda_2$ (attracting turbulent branch, solid line) as a function of $\epsilon^{-1}$ (which is proportional to the Reynolds number), for fiducial parameters $C_1 = 0.41$ and $C_2 = 0.54$ (see Section \ref{sec:parfit}), and for various values of the stratification parameter $\sigma$ (the value of $\sigma$ is marked near its corresponding curve). Right: Plot of $\epsilon \sigma_c(\epsilon)$ as a function of $\epsilon^{-1}$ for $C_1 = 0.41$ and $C_2 = 0.54$. This quantity is related to the critical value of $J {\rm Pr}$ in diffusive stratified shear flows (see text for detail).}
					\label{fig:bifur}
				\end{center}
			\end{figure}

			For instance, the unstratified limit ($\sigma = 0$) was studied by \citet{GaraudOgilvie2005}. In that case, 
			\begin{equation} 
			\frac{2C_2}{3} \lambda  = (C_1 \lambda  + \epsilon)  \left((C_1 + C_2)\lambda  + \epsilon\right)^2 ,
			\end{equation}
			whose solutions can be computed numerically for given values of $\epsilon$, $C_1$ and $C_2$ and have the following asymptotic approximation in the limit of small $\epsilon$ (or equivalently, when ${\rm Re}_S \rightarrow \infty$):  
			\begin{eqnarray}
			\lambda_1 =  \frac{3 \epsilon^3}{2 C^2 } + O(\epsilon^4), \label{eq:u1local}  \\ 
			\lambda_2 = \left( \frac{2C_2}{3C_1} \frac{1}{(C_1+C_2)^2} \right)^{1/2} - \frac{ 3C_1+ C_2}{C_1(C_1+C_2)} \epsilon + O(\epsilon^2) ,  \\
			\lambda_3 = - \left( \frac{2C_2}{3C_1} \frac{1}{(C_1+C_2)^2} \right)^{1/2} - \frac{ 3C_1+ C_2}{C_1(C_1+C_2)} \epsilon + O(\epsilon^2)  .
			\end{eqnarray}
			The negative solution $\lambda_3$ can be discarded on the basis that $\lambda$ has to be positive. As shown by \citet{GaraudOgilvie2005}, $\lambda_1$ corresponds to a repelling stationary turbulent state (and can therefore also be discarded) while $\lambda_2$ corresponds to the attracting stationary turbulent state that we seek. The solutions are shown in Figure \ref{fig:bifur}a, for for fiducial parameters $C_1 = 0.41$ and $C_2 = 0.54$. %This implies that the role of the solution $\lambda_1$ is merely to set the size of the basin of attraction of the turbulent solution, but is not otherwise relevant to the long-term statistically stationary dynamics of the turbulence either.  
			
			In the case where both $\sigma = 0$ and $\epsilon = 0$ the stress tensor components corresponding to the root $\lambda_2$ have simple exact analytical expressions \citep{GaraudOgilvie2005}:
			\begin{eqnarray}
			\lambda = \frac{\bar R^{1/2}}{L}  = \frac{1}{C_1+C_2} \sqrt{\frac{2C_2}{3C_1}} ,\label{eq:simplelambda} \nonumber \\
			r_{xz} = -  \frac{sC_1}{2(C_1+C_2)} \sqrt{\frac{2C_2}{3C_1}}  , \quad r_{xx} = \frac{3C_1 + C_2}{3(C_1+C_2)}, \nonumber \\
			r_{yy} = r_{zz} = \frac{C_2}{3(C_1+C_2)} .
			\label{eq:localresults}
			\end{eqnarray}
			Notably, we see that all of the ratios $r_{ij} = \bar R_{ij}/\bar R$ depend only on the ratio of the model constants $C_2/C_1$. This is an important property of the closure model that can both be tested numerically, and used to constrain $C_2/C_1$ (see Section \ref{sec:parfit}). 
			
			%Fig. \ref{fig:bifur} illustrates the behaviour of $\lambda_1$ and $\lambda_2$ as a function of $\epsilon^{-1}$ (which is proportional to the Reynolds number), for our fiducial parameters $C_1 = 0.4$ and $C_2 = 0.6$ (\citet{GO2005}). As pointed out by \citet{GO2005}, we recover the well-known result that the transition to turbulence in homogeneous unstratified shear flows occurs through a subcritical (finite amplitude) instability of the laminar solution.

			When $\sigma \neq 0$, equation (\ref{eq:u1local}) can only be solved numerically. Three branches of solutions exist, and as before, only two of them have $\lambda > 0$ and only one corresponds to an attracting state. They are shown in Fig. \ref{fig:bifur}a for different values of $\sigma $, as a function of $\epsilon^{-1}$ (which is proportional to the Reynolds number of the flow). We see that the transition to turbulence is subcritical at $\sigma = 0$ \citep{GaraudOgilvie2005} and remains subcritical as $\sigma$ increases. We also see that the critical Reynolds number above which turbulent solutions exist increases with $\sigma$. At fixed values of $\epsilon$, on the other hand, turbulent solutions disappear when $\sigma$ exceeds a critical value $\sigma_c(\epsilon)$, a result that is consistent with the notion that a shear flow can be stabilized by sufficiently large stratification even in the LPN limit. 
			
			In fact, it is interesting to note that 
			\begin{equation}
			\epsilon \sigma = \frac{C_{\nu}}{C_{\nu\kappa}} J {\rm Pr} ,
			\end{equation}			
			so this product is important for several reasons. First, it no longer depends on the (so far arbitrary) selection of the eddy scale $L$ (that both $\sigma$ and $\epsilon$ depend on), and secondly, the quantity $J {\rm Pr}$ was found to be the relevant bifurcation parameter for stratified shear flows in the LPN limit, with flows having $J {\rm Pr} < (J {\rm Pr})_c  \simeq 0.007$ being unstable, and flows with $J {\rm Pr} > (J {\rm Pr})_c $ being stable to diffusive shear instabilities (see Section \ref{sec:intro}). For this reason, we show in Figure \ref{fig:bifur}b the product $\epsilon \sigma_c(\epsilon)$ computed from the numerical solutions shown in Figure \ref{fig:bifur}a. We see that this product is not constant but instead, exhibits a well-defined maximum at intermediate values of $\epsilon^{-1}$ and tends to zero in the inviscid limit. The fact that $\epsilon \sigma_c(\epsilon) \rightarrow 0$ in the limit of small $\epsilon^{-1}$ is not actually surprising: a shear flow should indeed become stable even in the unstratified limit for sufficiently large viscosity. On the other hand, the fact that $\epsilon \sigma_c(\epsilon) \rightarrow 0$ in the limit of small viscosity (large $\epsilon^{-1}$) is somewhat more puzzling, since we would indeed expect that this quantity should tend to a constant in the inviscid limit, at least in the light Zahn's model \citep{Zahn1974}. As we shall demonstrate in Section \ref{sec:parfit}, however, this apparent discrepancy is not actually problematic, and some elements of Zahn's model can indeed be recovered. 			
			
			Finally the strictly inviscid limit $\epsilon = 0$ for arbitrary values of $\sigma$ is easy to solve analytically, since we have :
			\begin{equation} 
			\frac{2C_2}{3} \lambda =  \lambda \left[  4\sigma   \left(C_1 + \frac{C_2}{3} \right)   + C_1  (C_1 + C_2)\lambda  \right] \left[ 2\sigma + (C_1+C_2)\lambda  \right] .
			\end{equation}
			One of the roots is $\lambda_1= 0$, while the other two satisfy the quadratic
			\begin{equation} 
			\frac{2C_2}{3} =   \left[  4\sigma  \left(C_1 + \frac{C_2}{3} \right)   + C_1  (C_1 + C_2)\lambda  \right] \left[ 2\sigma+ (C_1+C_2)\lambda  \right] ,
			\label{eq:quadraticforlambda}
			\end{equation}
			whose solutions are
			\begin{equation}
			\lambda_{2,3} = -\left[  \left( 3 + \frac{2C_2}{3C_1} \right) \pm \sqrt{ \left( 1+ \frac{2C_2}{3C_1}\right)^2 + \frac{2C_2}{3C_1} \frac{1 }{\sigma^2} }  \right] \frac{\sigma}{C_1+ C_2} .
			\end{equation}
			%Checked 
			While this always has real solutions, a necessary condition for at least one of these solutions to be positive is : 
			\begin{equation}
			4 \left( \frac{3C_1}{C_2}  + 1\right) \sigma^2  \le 1  .
			%Checked 
			\end{equation}
			This states that turbulence dominated by eddies of scale $L$ can be sustained in inviscid low P\'eclet number stratified shear flows provided $ \sigma <  \sigma_c(\epsilon=0) = 0.5(3C_1/C_2+1)^{-1/2} \simeq 0.276$ for the fiducial parameters $C_1 = 0.41$, $C_2 = 0.54$ (see Section \ref{sec:parfit}). The disappearance of the turbulent solution as $\sigma$ approaches this critical value can be clearly seen in Fig. \ref{fig:bifur}a. 
			Interestingly, if we define the P\'eclet number of eddies on scale $L$ as ${\rm Pe}_L  = L^2 {\rm Pe}_S$  then we can express the condition for instability in the inviscid limit as
			\begin{equation}
			J {\rm Pe}_L \le  \frac{C_{\nu\kappa}}{2} \left( \frac{3C_1}{C_2}  + 1\right)^{-1/2}  .
			\end{equation}
			%where $(J {\rm Pe})_c$ is a constant of order unity\footnote{The constant $(J {\rm Pe})_c$ was estimated to be of order 0.5 by \citet{Garaudal17} in simulations of stratified plane Couette flows. It is tempting to use this result to constrain $C_{\nu\kappa}$ given $C_1$ and $C_2$, but it is worth remembering that the estimate of \citet{Garaudal17} is inherently tied to the manner in which they define the turbulent eddy size, so $(J {\rm Pe})_c$ is not a universal constant, and must be determined on a case by case basis.}. 
			This criterion is reminiscent of the findings of \citet{Townsend58} and \citet{Dudis1974} for the linear instability of stratified shear flows in the optically thin regime, even though it is applied here to optically thick homogeneous stratified shear flows that are linearly stable \citep{Knobloch84} but nonlinearly unstable. It is also at the heart of Zahn's general nonlinear criterion for diffusive shear instabilities (see Section \ref{sec:intro} and Paper I for a review on the topic). It is therefore reassuring to see that our closure model captures this essential feature of diffusive shear instabilities. 
			
\begin{figure}[!ht]
				\begin{center}
					\includegraphics[width=.5\textwidth]{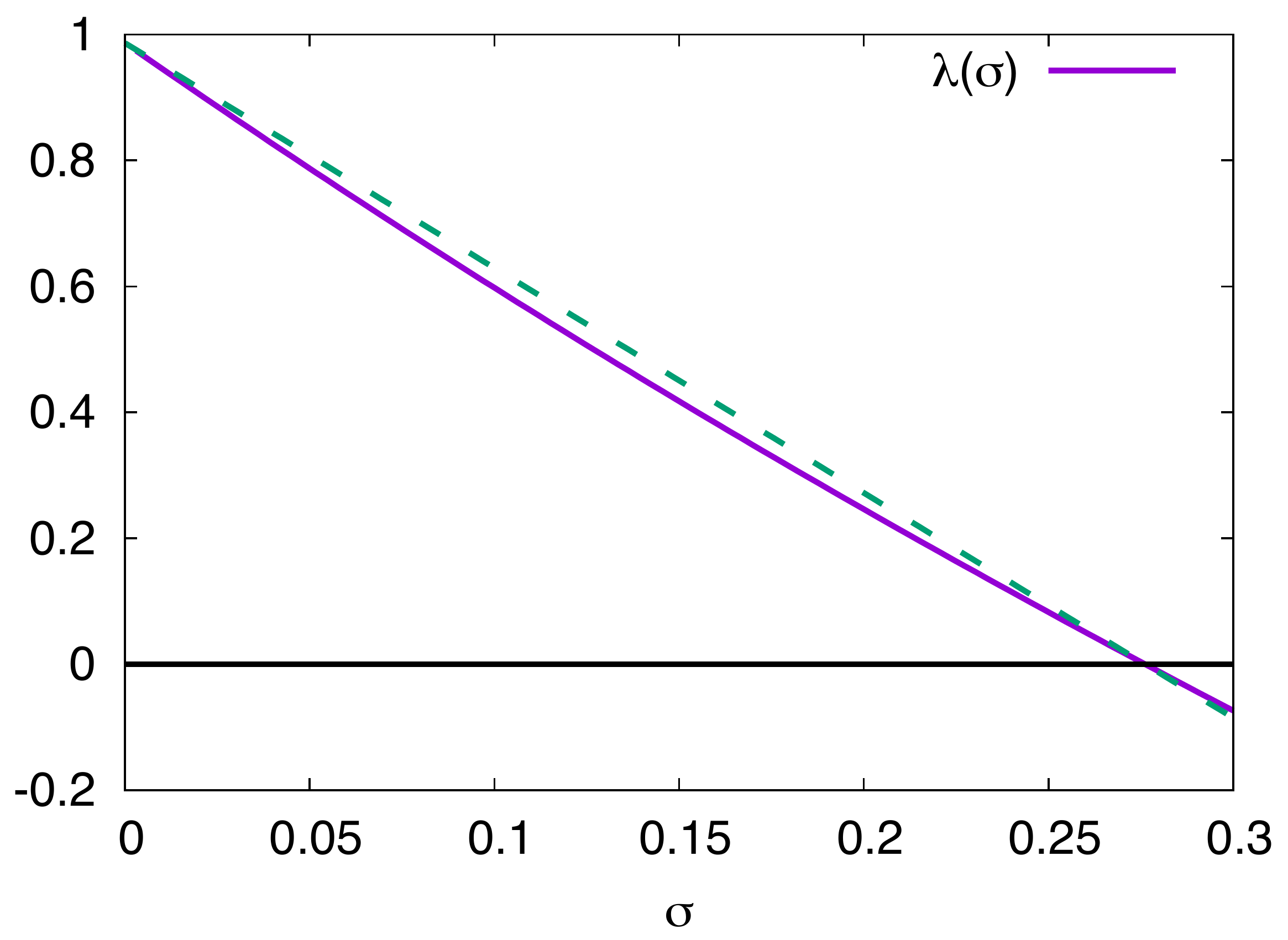}
					\caption{ Numerical solutions for $\lambda(\sigma)$ in the limit of $\epsilon \rightarrow 0$, for our parameters $C_1 = 0.41$ and $C_2 = 0.54$ (solid line). Also shown is the curve $ \sqrt{2C_2/3C_1} (1- \sigma/\sigma_c)/(C_1+C_2)$ (dashed line), which is a relatively good approximation to the true solution for these parameters.}
					\label{fig:quadraticforlambda}
				\end{center}
			\end{figure}

			The positive solution of (\ref{eq:quadraticforlambda}) is shown in Figure \ref{fig:quadraticforlambda} for $C_1 = 0.41$, $C_2 = 0.54$, and for $0 <\sigma < \sigma_c(\epsilon=0)$. It happens to be relatively well-approximated by a linear function, with $\lambda \simeq \sqrt{2C_2/3C_1} (1- \sigma/\sigma_c)/(C_1+C_2)$ for $0 \le \sigma \le \sigma_c$, and this formula can then be used to reconstruct all the components of the stress tensor if an approximate solution to the local closure model is all that is needed.
 			
			 \subsection{Comparison with numerical simulations of plane Couette flow and parameters fitting}
	\label{sec:parfit}

			The results of the local model can be used to constrain the closure parameters $C_1$, $C_2$, $C_\nu$ and $C_{\nu\kappa}$, and to gain better insight into possible strategies for selecting the eddy scale $L$. Ideally, we should constrain the parameters by comparing the model with simulations of homogeneous stratified shear flows, such as the ones produced by \citet{PratLignieres13,PratLignieres14} or \citet{Pratal2016}. However, these papers do not report the values of all the components of the stress tensor measured in their simulations, and can therefore not be used for our purpose. Instead, we elect to compare the model with the results of DNSs of stratified plane Couette flow of Paper I. While such a  setup was not originally designed to produce homogeneous stratified turbulence (because it is bounded by the top and bottom plates), we have found that many of the turbulence properties in the central region of the domain, far from the boundaries, are quantitatively consistent with those measured in the shearing box simulations of \citet{Pratal2016} (see Paper I for detail). In what follows, we first briefly summarize the DNSs of Paper I and explain how various quantities of interest such as the Reynolds stresses are extracted, and then use them to constrain the closure parameters. 
			
			\subsubsection{Description of the DNSs from Paper I}
		
			Paper I presents DNSs of stratified plane Couette flow in the LPN limit. This setup is boundary driven: (1) the shear is driven between two no-slip plane-parallel plates moving at velocities $\pm \Delta U / 2 $ respectively, and (2) a background temperature gradient is maintained by holding the top and bottom plates at temperatures $\pm \Delta T /2$. The vertical distance between the two plates is $L_z$, and the domain is assumed to be periodic in the horizontal direction. The system dynamics are evolved using the nondimensional LPN equation
		\begin{equation}
		\frac{\partial\bu}{\partial t} + \bu \cdot \bnabla \bu = -\bnabla p + \Ri_{\rm C}\Pe_{\rm C}\bnabla^{-2}u_z\be_z + \frac{1}{\Ree_{\rm C}} \nabla^2\bu  \mbox{  with   } \nabla \cdot \bu = 0 \label{eq:momentumLPNpaper1}, \\
		\end{equation} 
	        where $\Ri_{\rm C}  = \alpha g L_z \Delta T / \Delta U^2$, $\Pe_{\rm C} = \Delta U L_z / \kappa_T$  and $\Ree_{\rm C} = \Delta U L_z / \nu$ are the global Richardson, P\'eclet and Reynolds numbers for this problem. In all cases, the solution is evolved  numerically in time until a statistically stationary state is reached at which point the simulation is analyzed to extract the quantities of interest (see Paper I for details of the code). In that state, the shear in the bulk of the domain is nearly constant aside from two thin viscous boundary layers located in the immediate vicinity of each plate. Paper I already reported the value of the mean shear in the bulk of the domain (away from the boundaries) $S$ (reported as $\hat S$ and measured in units of $\Delta U / L_z$), the component of the Reynolds stress tensor $\bar R_{xz}$ (reported as $\langle uw \rangle_t$ and measured in units of $\Delta U^2$), as well as the typical eddy scale $l_e$ (in units of $L_z$) in Table 1. See Paper I for details on the method used to obtain these bulk quantities. We have re-analyzed the same simulations to extract in addition $\bar R_{xx}$, $\bar R_{yy}$, and $\bar R_{zz}$ in the bulk of the domain, for the two highest Reynolds number runs only (${\rm Re}_{\rm C} = 9 \times 10^4$ and $1.2 \times 10^5$). 
	        
	        Knowing these three components we can then calculate $\bar R = \bar R_{xx} + \bar R_{yy} + \bar R_{zz}$, and then the corresponding ratios $r_{xz}$, $r_{xx}$, $r_{yy}$ and $r_{zz}$ (see Section \ref{sec:local1}). Finally, the value of $J {\rm Pe}_S$ corresponding to the shear in the middle of the domain is calculated using the formula $J {\rm Pe}_S = \Ri_{\rm C} \Pe_{\rm C} / S$. 
			
			\subsubsection{Constraining the model parameters}
			
\begin{figure}[!ht]
				\begin{center}
					\includegraphics[width=.5\textwidth]{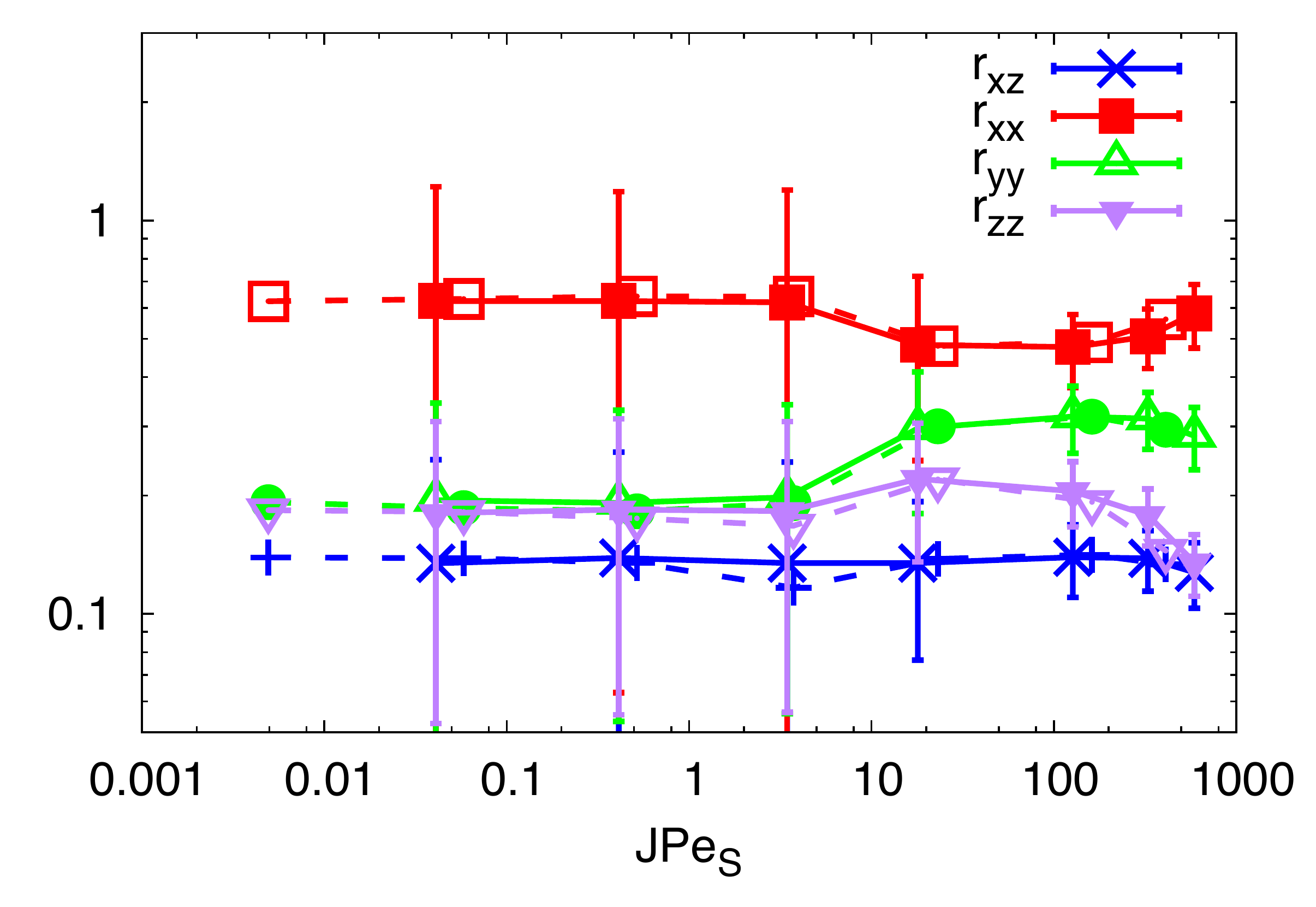}
					\caption{Variation of the ratios $r_{xz}$, $r_{xx}$, $r_{yy}$ and $r_{zz}$ with $J {\rm Pe}_S$ in the two highest Reynolds number sets of simulations presented in Paper I. The dashed lines correspond to the $\Ree_{\rm C} = 9\times 10^4$ simulations while the solid lines correspond to the $\Ree_{\rm C} = 1.2 \times 10^5$ simulations. The errorbars reflect the measured r.m.s. fluctuations of each of the $\bar R_{ij}$; only those for the $\Ree_{\rm C} = 1.2 \times 10^5$ are shown to avoid crowding the figure.}
					\label{fig:localstresstest}
				\end{center}
			\end{figure}

			We begin by comparing the results of DNSs in the very weakly stratified limit with the analytical formulas given in (\ref{eq:localresults}). As discussed in Section \ref{sec:local1}, the local model predictions for the ratios $r_{xz}$, $r_{xx}$, $r_{yy}$ and $r_{zz}$ are independent of the eddy scale $L$, and only depend on the ratio $C_2/C_1$, which provides a way of testing the model and constraining that ratio using several independent measurements. Figure \ref{fig:localstresstest} shows $r_{xz}$, $r_{xx}$, $r_{yy}$ and $r_{zz}$ as a function of $J{\rm Pe}_S$, for the two highest Reynolds number sets of simulations available. We see that the results of the two sets are consistent with one another, confirming that the Reynolds number is indeed sufficiently high to be in the inviscid limit for these runs. 
			
			As predicted by the local closure model, $r_{yy}$ is indeed very close to $r_{zz}$ in the limit of weak stratification ($J {\rm Pe}_S \ll 1$). We also see that in the same limit, $r_{xx} \simeq 0.62$, $r_{zz} \simeq 0.19$ and $r_{xz} \simeq 0.15$. Comparing these values with their analytical predictions would imply that $C_2/C_1$ should be roughly equal to $1.32$, $1.32$ and $1.34$ respectively, three values that are remarkably consistent with one another. This is an encouraging indication that the closure model is behaving appropriately in this unstratified limit. In order to constrain $C_1$ and $C_2$ individually, we use the results of \citet{GaraudOgilvie2005} who noted that the quantity 
			\begin{equation}
			\kappa = \frac{2}{C_1} \left[ \frac{C_1 C_2}{6(C_1+C_2)^2} \right]^{3/4}
			\end{equation}
			is the well-known von K\'arm\'an constant of the Prantl-von K\'arm\'an universal velocity profile for wall-bounded shear flows. This constant has been measured in a number of laboratory experiments, notably those of \citet{ZagarolaSmits1998}, and found to be $\kappa = 0.436$. Using $C_2/C_1 = 1.32$ in this formula, we finally find that  $C_1 \simeq 0.41$ and so $C_2 \simeq  0.54$. The value of $C_1$ found is remarkably close to that originally proposed by \citet{GaraudOgilvie2005}, while $C_2$ is smaller than in \citet{GaraudOgilvie2005} by 10 percent only. In what follows, we therefore adopt $C_1 = 0.41$ and $C_2 = 0.54$ as our fiducial parameters. 

\begin{figure}[!ht]
				\begin{center}
					\includegraphics[width=.5\textwidth]{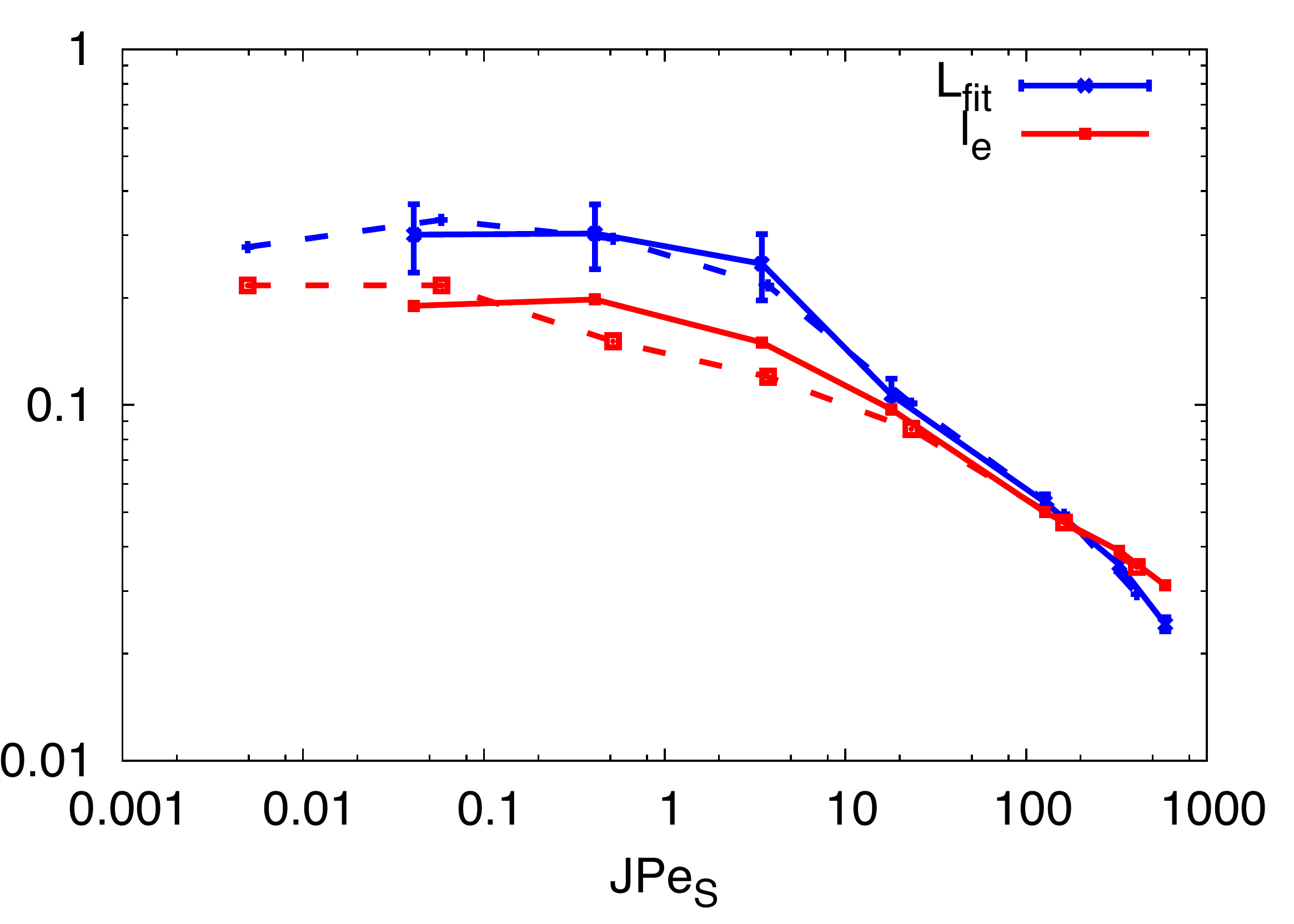}
					\caption{Comparison of the value of $L_{\rm fit}$ required to fit the data from the DNSs of Paper I for $\bar R^{1/2}$ (blue), with the measured eddy scale $l_e$ (red). The dashed lines corresponds to the $\Ree_{\rm C} = 9\times 10^4$ simulations while the solid lines correspond to the $\Ree_{\rm C} = 1.2 \times 10^5$ simulations. The errorbars on $L_{\rm fit}$ reflect the measured r.m.s. fluctuations of $\bar R$; only those for the $\Ree_{\rm C} = 1.2 \times 10^5$ are shown to avoid crowding the figure. Errorbars on $l_e$ are not easy to measure, but would be of the same order as those measured for $L_{\rm fit}$. Note that the comparison between $l_e$ and $L_{\rm fit}$ is only meaningful in the limit $J {\rm Pe}_S \ll 1$, where equation (\ref{eq:simplelambda}) applies. We find that in this limit $L_{\rm fit} \simeq 1.5 l_e$.  }
					\label{fig:localltest}
				\end{center}
			\end{figure}

			One would also like to compare the model prediction for $\bar R$ (given in equation \ref{eq:simplelambda}) with the available data. Unlike the ratios $r_{ij}$, however, $\bar R$ depends on the assumed eddy-scale $L$ that has so far remained unspecified. For this reason, we use a different approach: instead of picking an arbitrary model for $L$ and comparing the predicted values of $\bar R$ with the data, we use equation (\ref{eq:simplelambda}) to compute what value of $L$ would be required to reproduce the values of $\bar R$ measured in the DNSs for low stratification. In other words, we compute the quantity $L_{\rm fit} = (C_1 + C_2) (3C_1/2C_2)^{1/2} \bar R^{1/2}$ from the simulations, using the  fiducial values of $C_1$ and $C_2$ selected earlier. The results are shown in Figure \ref{fig:localltest} for the two sets of DNSs from Paper I with the highest Reynolds numbers, and compared with the actual eddy scale $l_e$ measured from the same DNSs (see Table 1 of Paper I). Note that the comparison is only meaningful in the limit where $J {\rm Pe}_S \ll 1$, for which (\ref{eq:simplelambda})  applies. We see that the value $L_{\rm fit}$ required to fit the $\bar R$ data is a little larger than $l_e$ by a factor $1.5$ in this limit. The discrepancy is not surprising, given that the meanings of $L$ and $l_e$ are very different: one is created from an energy cascade timescale in a closure model, and the other is measured from DNSs using an autocorrelation lengthscale (see Paper I). As such, we do not expect them to be identical a priori, but we do expect them to remain proportional to one another as parameters are varied. In the remainder of this section, and for the purpose of calibrating the remaining model parameters $C_\nu$ and $C_{\kappa\nu}$ we now take $L = 1.5 l_e$ where $l_e$ is the measured eddy scale from DNSs. This assumption is successfully verified in Section \ref{sec:Kolmo}.

\begin{figure}[!ht]
				\begin{center}
					\includegraphics[width=.5\textwidth]{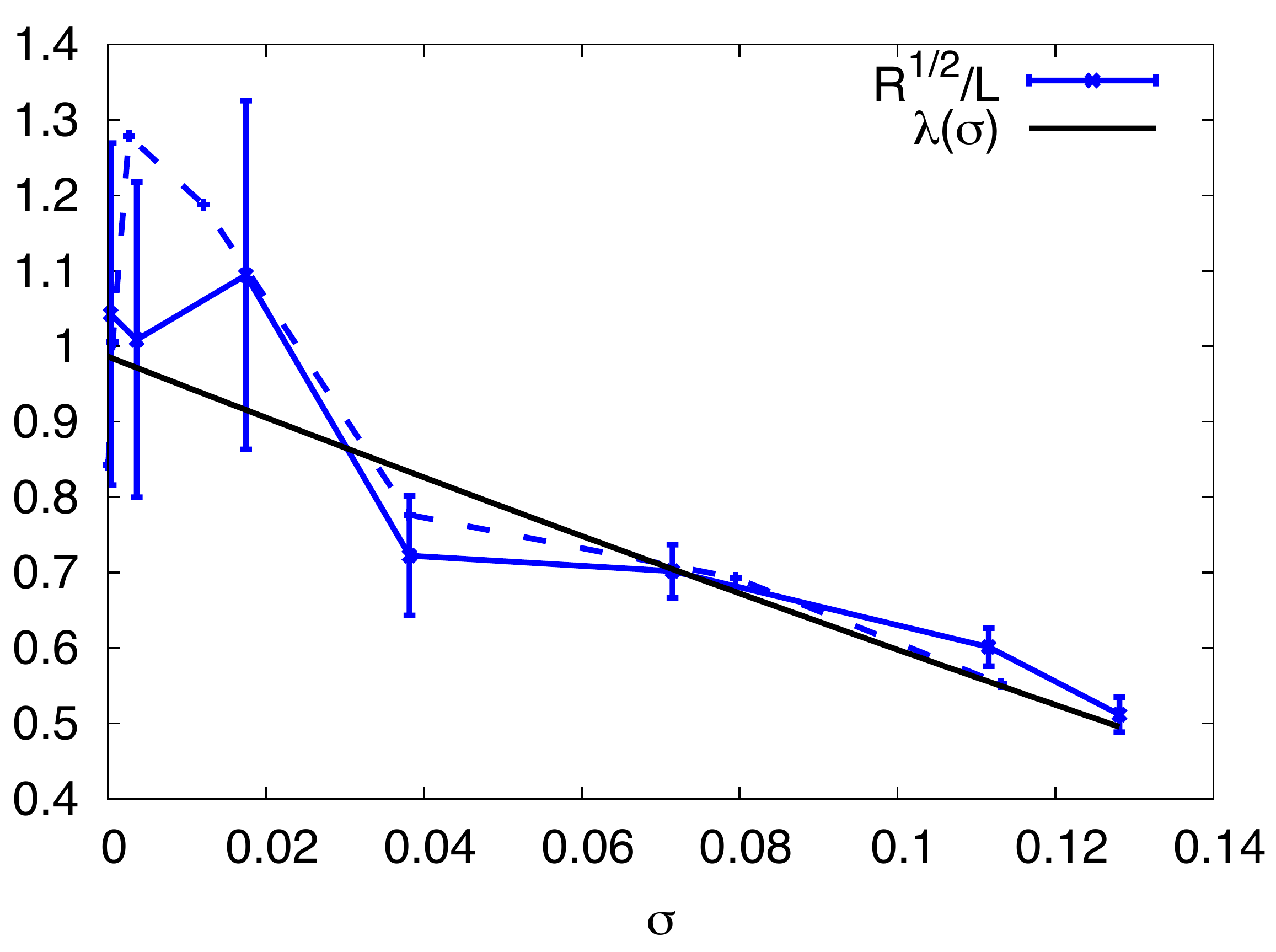}
					\caption{Comparison of the left-hand-side and right-hand-side of equation (\ref{eq:fitcnuk}), using the data from the DNSs of Paper I to compute $\bar R^{1/2} / L$, with $L$ taken to be $1.5 l_e$. The dashed lines corresponds to the $\Ree_{\rm C} = 9\times 10^4$ simulations while the solid lines correspond to the $\Ree_{\rm C} = 1.2 \times 10^5$ simulations. The errorbars on $\bar R^{1/2}/L$ reflect the measured r.m.s. fluctuations of $\bar R$; only those for the $\Ree_{\rm C} = 1.2 \times 10^5$ are shown to avoid crowding the figure. The best fit, shown here, is obtained with $C_{\nu\kappa} = 10$.}
					\label{fig:localCnukappatest}
				\end{center}
			\end{figure}

	In order to constrain $C_{\nu\kappa}$, we must now look at the limit where $\sigma$ is not asymptotically small. Assuming as before that viscosity is negligible, we have 
	\begin{equation}
	\frac{\bar R^{1/2}}{L} = \lambda(\sigma) = -\left[  \left( 3 + \frac{2C_2}{3C_1} \right) \pm \sqrt{ \left( 1+ \frac{2C_2}{3C_1}\right)^2 + \frac{2C_2}{3C_1} \frac{1 }{\sigma^2} }  \right] \frac{\sigma }{C_1+ C_2} .\label{eq:fitcnuk}
        \end{equation}
        The right-hand side of this expression is the function $\lambda(\sigma)$ shown in Figure \ref{fig:quadraticforlambda}, which only depends on $\sigma = J {\rm Pe}_S L^2 / C_{\nu \kappa}$, where $ J {\rm Pe}_S L^2 $ is known from the data (assuming that $L = 1.5l_e$ where $l_e$ is measured in the DNSs). This implies that it contains only one unknown, namely $C_{\nu\kappa}$. Meanwhile the left-hand side can be directly measured from the available data too. We can then vary $C_{\nu\kappa}$ (which varies $\sigma$) to find the best fit between the left- and right-hand-sides of equation (\ref{eq:fitcnuk}). The results are shown in Figure \ref{fig:localCnukappatest}, with a best fit obtained for $C_{\nu\kappa} \simeq 10$. This is fairly significantly larger than the value of $C_{\nu\kappa} \simeq 6$ estimated by \citet{Garaudal2010} by fitting their related closure model against DNSs of Rayleigh-B\'enard convection but still within a factor of two of the latter. %It is worth noting that with this selected value of $C_{\nu\kappa}$, the DNSs appear to be limited to $\sigma \le 0.13$, even though the local closure model itself supports inviscid solutions up to $\sigma_c \sim 0.276$ with the selected values of $C_1$ and $C_2$. 
        
	The most difficult parameter to constrain is $C_\nu$, since the latter is only relevant when viscosity becomes relevant, a limit we have so far avoided. Comparing the model to the lowest Reynolds number DNSs of Paper I is not a good idea, because the influence of the viscous boundary layers (where the local model breaks down) is larger for lower Reynolds numbers. Instead, we calibrate $C_\nu$ (which appears in the definition of $\epsilon)$ to ensure that the closure model correctly predicts the critical value of the stratification above which the shear flow becomes stable in the DNSs. 
	
\begin{figure}[!ht]
				\begin{center}
					\includegraphics[width=.5\textwidth]{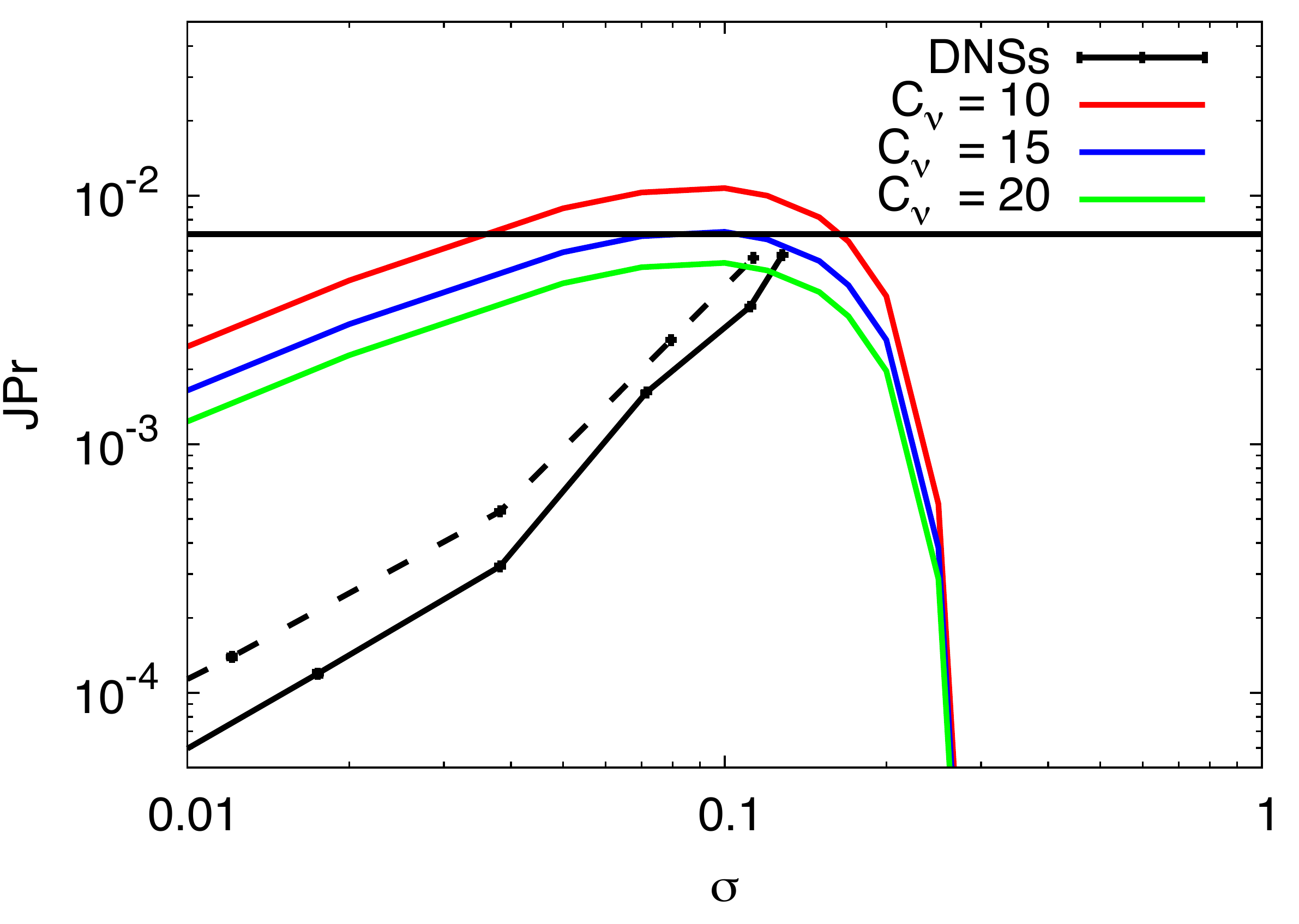}
					\caption{This figure shows the values of $J$Pr measured in the DNSs of Paper I (with the solid line corresponding to the $\Ree_{\rm C} = 9\times 10^4$ simulations while the dashed lines correspond to the $\Ree_{\rm C} = 1.2 \times 10^5$ simulations), and compares this with the various predicted critical threshold for instability: the horizontal black line shows the simple model given by equation (\ref{eq:Zahncrit}) with $(J{\rm Pr})_c = 0.007$; the colored lines show the predicted values of the critical threshold for stability in our closure model, namely $\sigma \epsilon_c(\sigma) C_{\nu\kappa}/C_\nu$, for fiducial values of the parameters calibrated so far ($C_1 = 0.41$, $C_2 = 0.54$ and $C_{\nu\kappa} = 10$), for different values of $C_\nu$. The curve with $C_\nu  =15$ appears to be the better choice in comparison with the DNSs. }
					\label{fig:localCnutest}
				\end{center}
			\end{figure}

	As explained by \citet{Zahn1974} and seen in Figure \ref{fig:localltest}, the dominant eddy scale of diffusive shear instabilities decreases as the stratification increases (see Paper I for more on this topic). For sufficiently large stratification, that scale becomes viscously damped, at which point the shear is completely stabilized. In Paper I, we showed that this happens in DNSs when $J {\rm Pr}= (J {\rm Pr})_c \simeq 0.007$, while from the closure model properties we find that this happens when $\sigma \sim \sigma_c(\epsilon)$.  Earlier, we also showed that $\epsilon \sigma = J {\rm Pr} C_\nu/ C_{\nu\kappa}$. As a result, we can calibrate $C_\nu$ to ensure that the numerically determined stability limit $J {\rm Pr} \simeq 0.007$ is consistent with the closure model limit $\sigma = \sigma_c(\epsilon)$ (or equivalently, $\epsilon \sigma = J {\rm Pr} C_\nu/ C_{\nu\kappa} = \epsilon \sigma_c(\epsilon)$). This is done in Figure \ref{fig:localCnutest}, which shows $J {\rm Pr}$ as a function of $\sigma$, and compares several curves: the numerically-determined threshold $ (J {\rm Pr})_c $ from Paper I, the actual data points from the two DNSs with the highest Reynolds numbers of Paper I, and the critical threshold for stability in our closure model, namely $\sigma \epsilon_c(\sigma) C_{\nu\kappa}/C_\nu$, for $C_{\nu\kappa} = 10$ and different values of $C_\nu$. The value of $C_\nu$ which appropriately captures the correct threshold from the numerical simulations is $C_\nu \sim 15$. This value is very much consistent with the one originally estimated by \citet{GaraudOgilvie2005}, namely $C_\nu \sim 12.5$, which is remarkable given the fact that the latter was determined using a completely different approach (by fitting the velocity profile for wall-bounded shear flows).
	
	To conclude, in what follows with therefore adopt the parameters $C_1 = 0.41$, $C_2 = 0.54$, $C_{\nu\kappa} = 10$ and $C_\nu = 15$. Note that we have refrained from providing errorbars on the estimated values for the model parameters, because these errorbars would only be meaningful in the context of the specific method we have used to fit them. In other words, had we selected a different method or a different dataset, we would likely have obtained somewhat different values of the parameters, with their own errorbars. The only meaningful way to estimate the errors on $C_1$, $C_2$, $C_{\nu\kappa}$ and $C_\nu$ is to fit the model to many different types of systems, and see how different the estimated parameters are. We have already seen that the values of $C_1$, $C_2$ and $C_\nu$ fitted against the Paper I DNSs are within $\sim 20\%$ of the corresponding values measured by \citet{GaraudOgilvie2005}. Meanwhile $C_{\nu\kappa}$ is within $\sim 50\%$ of the value proposed by \citet{Garaudal2010}. Hence, 20\% (respectively 50\%) could be a good estimate of the size of the error on the model parameters $C_1$, $C_2$, $C_\nu$ (respectively $C_{\nu\kappa}$). 
       
	\section{Comparison of the closure model predictions with DNSs of stratified Kolmogorov flows in a statistically-stationary state}
	\label{sec:Kolmo}

	Having estimated the closure model parameters $C_1$, $C_2$, $C_\nu$ and $C_{\nu\kappa}$ from DNSs of stratified plane Couette flow, we now compare the predictions of the model with results from DNSs of stratified Kolmogorov flows (i.e. spatially sinusoidal flows) of \citet{GaraudKulen16}. These sets of simulations solve the LPN equations 
	\begin{equation}
		\frac{\partial\bu}{\partial t} + \bu \cdot \bnabla \bu = -\bnabla p + \Ri_F\Pe_F\bnabla^{-2}u_z\be_z + \frac{1}{\Ree_F} \nabla^2\bu  + \hat F(z) \be_x  \mbox{  with   } \nabla \cdot \bu = 0  \label{eq:momentumLPNpaper2}, \\
		\end{equation} 
		where the non-dimensionalization used is based on the amplitude $F_0$ and wavenumber $k$ of the sinusoidal forcing function \citep[see][for detail]{GaraudKulen16}. The unit length $L_0 = k^{-1}$ and velocity $U_0 \equiv U_F = (F_0/k \rho_0)^{1/2}$ are used to create appropriate Richardson, P\'eclet and Reynolds numbers as $\Ri_F = N^2 / k^2 U_F^2$, $\Pe_F = U_F / k \kappa_T$  and $\Ree_F = U_F / k \nu$. The non-dimensional forcing becomes $\hat F(z) = \sin(z)$, and the non-dimensional height of the domain is $2\pi$. The perturbations are triply periodic. All runs available are summarized in Table 3 of \citet{GaraudKulen16}. In what follows, we focus once again on the highest Reynolds number simulations only (which have $\Ree_F = 100$). 		
		
		From each of the runs, we re-analyzed the data to collect instantaneous $z-$dependent profiles of the horizontally averaged flow $\bar u_x(z,t)$ and Reynolds stresses $\bar R_{ij}(z,t)$ once the system has reached a statistically-stationary state. These can then be directly compared with the closure model predictions for given values of the input parameters, and of the assumed eddy scale $L$.
		
		In what follows, we use the fiducial values of the closure parameters obtained in Section \ref{sec:local}. We then assume that $L$ is constant in the domain, and find the value of $L$ that best fits a particular simulation (i.e. for a given value of $\Ri_F \Pe_F$ at $\Ree_F = 100$). We use a nonlinear least-square Levenberg-Marquardt method to iteratively select the value of $L$ that minimizes the distance between the time-average of the mean flow profile $\bar u_x(z,t)$ obtained from the DNSs, and the model prediction for $\bar u_x(z)$. At each iteration, we make sure to evolve the closure model equations (i.e. equations (\ref{eq:dcm1})-(\ref{eq:dcm4}) as well as (\ref{eq:dcmt1}) and  (\ref{eq:dcmt3}), together with the mean flow equation (\ref{eq:cm2}) with $\bar T = 0$) in time until a steady state is reached\footnote{We attempted to use the nonlinear least-square method directly on the steady-state model equations, but were not able to find converged solutions in that case, presumably due to the nonlinearity of the closure model equations.}. 
		
\begin{figure}[!ht]
				\begin{center}
					\includegraphics[width=.5\textwidth]{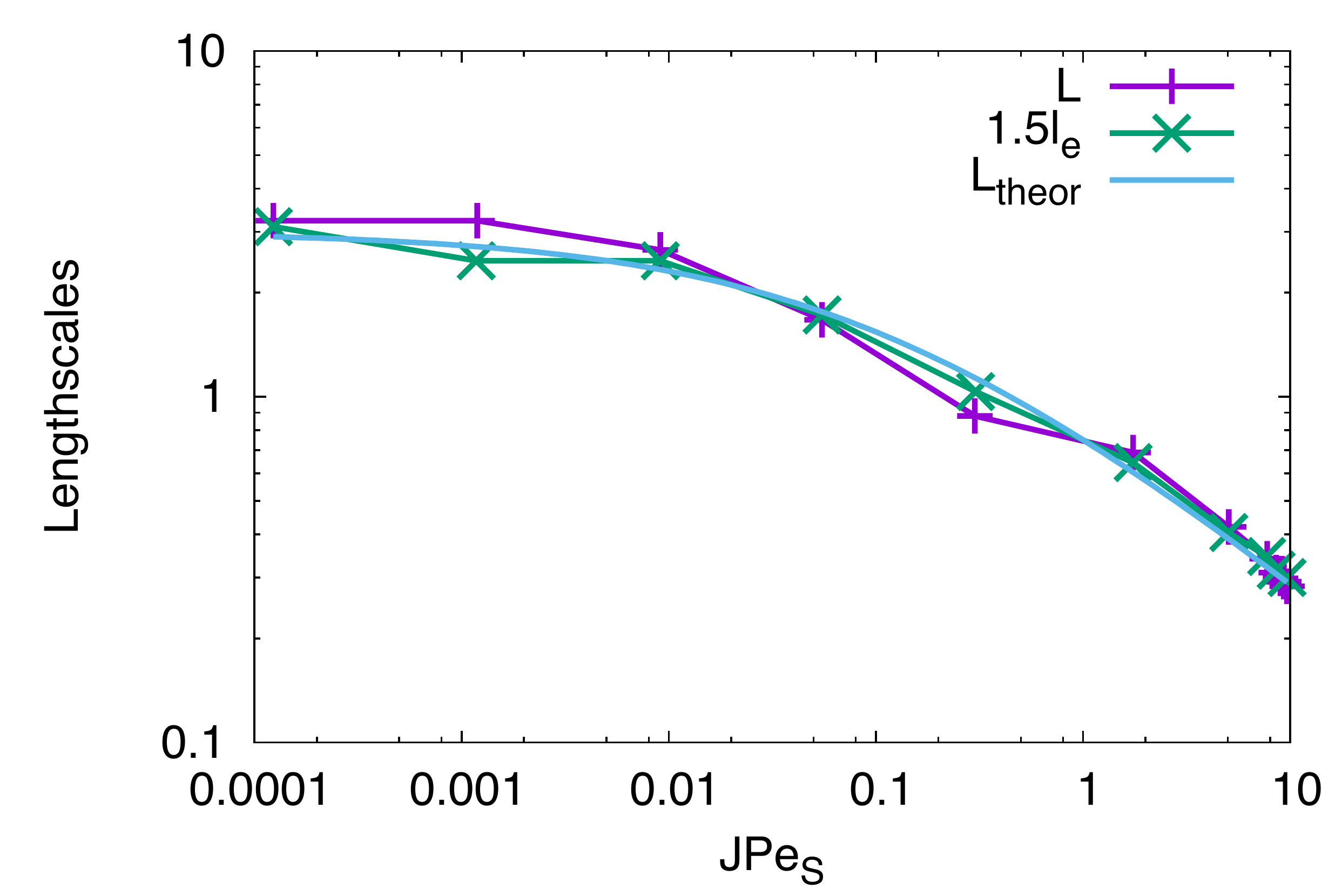}
					\caption{Comparison between the best fit for $L$, the measured eddy scale $l_e$ times 1.5, and the model $L_{\rm theor}$ (see equation (\ref{eq:Ltheor}) in Section \ref{sec:tanh}), as a function of $J \Pe_S$. }
					\label{fig:Lcompare}
				\end{center}
			\end{figure}

		The values of $L$ thus obtained, for which the closure model is best able to predict the mean flow, are presented in Figure \ref{fig:Lcompare} for each simulation, as a function of $J \Pe_S = \Ri_F \Pe_F / |S|$ where $S$ is the shearing rate measured in the middle of the domain (where it is largest). We see that as expected, $L$ is commensurate with the size of the shear layer for weakly stratified flows, and decreases progressively as $J \Pe_S$ increases. We also measured the turbulent eddy scale from the DNSs directly, using the first zero vertical autocorrelation function of the spanwise ($y-$) velocity, as in equation (33) of Paper I. The latter is also shown in Figure \ref{fig:Lcompare}. Remarkably, we see that as assumed in Section \ref{sec:local}, $L$ is very well predicted by $1.5 l_e$ for {\it all} available simulations. This firmly establishes that the quantity $L$ introduced in the closure model merely as a means to compute a timescale for the turbulent cascade really has a genuine physical counterpart as the vertical size of the turbulent eddies. It also implies that, moving forward, creating a good model for $L$ to be used in the closure is equivalent to creating a good model for the actual eddy scale $l_e$. 	This will be revisited in Section \ref{sec:L}.

	       Let us now look in more detail at the comparison between the closure model prediction and the DNS results, for each component of the Reynolds stress tensor. We focus on three regimes: weak ($\Ri_F\Pe_F = 0.01$), intermediate ($\Ri_F \Pe_F = 1$) and strong ($\Ri_F \Pe_F = 100$) stratification.

		\subsection{The nearly unstratified limit}

\begin{figure}[!ht]
			\begin{center}
				\includegraphics[width=0.8\textwidth]{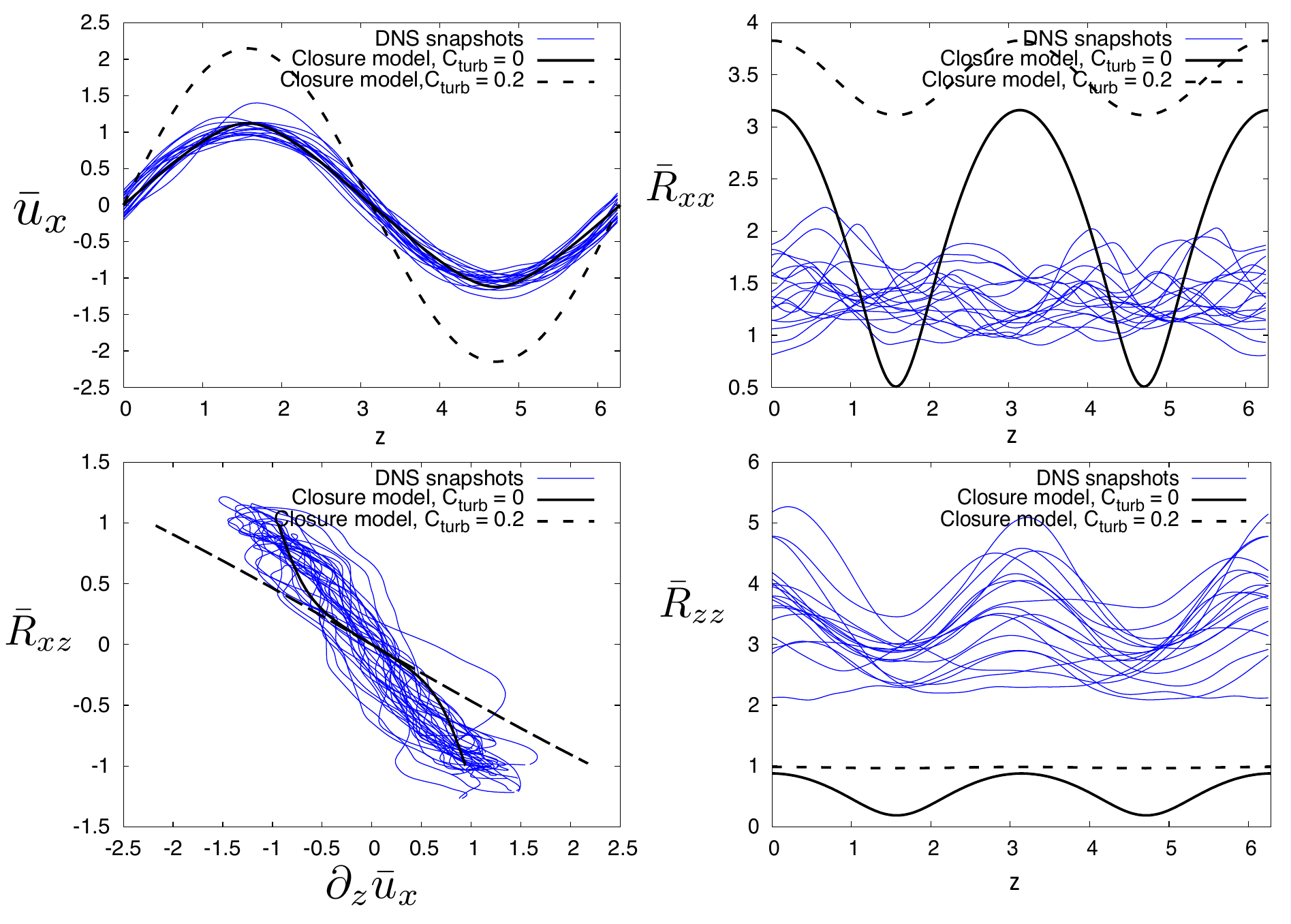}
				\caption{Comparison between snapshots of the horizontally averaged horizontal flow $\bar u_x(z)$, and of the Reynolds stress components $\bar R_{xz} ( \partial_z \bar u_x) $, $\bar R_{xx}(z)$ and $\bar R_{zz}(z)$, with steady state predictions from the closure model for $\Ree_F = 100$, $\Ri_F \Pe_F = 0.01$, and $L = 2.7$, together with fiducial model parameters. The case with $C_{\rm turb} = 0$ corresponds to the model discussed prior to Section \ref{sec:Cturb}, while the case with $C_{\rm turb} = 0.2$ corresponds to the model discussed in Section \ref{sec:Cturb}.}
				\label{fig:closure10.01}
			\end{center}
		\end{figure}

		Fig. \ref{fig:closure10.01} shows in blue snapshots in time of $\bar u_x(z)$, and of the Reynolds stress components $\bar R_{xz}(\partial_z \bar u_x) $, $\bar R_{xx}(z)$ and $\bar R_{zz}(z)$,  once the system has reached a statistically stationary state, from a DNS with $\Ree_F=100$ and $\Ri_F\Pe_F=0.01$ from \citet{GaraudKulen16}. Also shown in the solid black line are the corresponding predictions from the closure model, with fiducial parameters, for $L = 2.7$ (which provides the best fit for this simulation).   

		As reported by \citet{GaraudKulen16}, these low stratification DNSs favor large-scale perturbations that are highly variable in time. The mean flow is nevertheless close to being sinusoidal and in phase with the forcing, which suggests that the turbulence in this limit simply acts as a turbulent diffusivity. We also see that $\bar R_{zz}$ is minimal in regions of low shear and maximal in region of strong shear, showing that the perturbations are shear driven, as expected. However, the fact that these quantities are non-zero in regions where the shear is exactly null (here, at $z=\pi/2$ and $z = 3\pi/2$) also shows that there is significant turbulent overshoot from the adjacent sheared regions (also see Paper II). 
			
	      The value of $L$ selected ensures that the closure model best fits the measured mean flow profile, so it is not surprising to find that the two are in very good agreement. The agreement between the model and the data for $\bar R_{xz}(z)$ (not shown) is similarly good, which is not surprising since $\bar R_{xz}(z)$ in this body-forced setup is strongly constrained independently of any other quantity. Indeed starting from the mean streamwise equation in a steady state we have
	      \begin{eqnarray}
		\partial_z\bar R_{xz} & =  &  \frac{1}{\Ree_F} \partial_{zz} \bar u_x + \sin(z)\label{eq:umeaneq2}.
		\end{eqnarray}
		In the limit of large Reynolds number, the diffusive term can be neglected leaving a balance between momentum transport by the Reynolds stresses and the imposed forcing. This balance can be integrated straightforwardly to yield
		\begin{eqnarray}
		\bar R_{xz} & \simeq & - \cos(z)\label{eq:umeaneq3},
		\end{eqnarray}
so it is not surprising that this is found to be the case both for the data and for the closure model. On the other hand, a more sensitive test of the quality of the closure model consists in plotting $\bar R_{xz}$ against the local shearing rate $\partial_z \bar u_x$. We see that in this weakly stratified system, the two are in relatively good agreement, with the closure model predictions within the limits of variability of the DNSs. However, the model exhibits a somewhat more pronounced $s-$shaped curve than the DNSs (for which $\bar R_{xz}$ is very close to being a linear function of $\partial_z \bar u_x$). Finally, the agreement between the predicted stationary state profiles for $\bar R_{xx}$, $\bar R_{yy}$ (not shown) and $\bar R_{zz}$ and the data is fairly poor. $\bar R_{zz}$ is significantly under-estimated by the model for all $z$. The model also predicts a much stronger variation of $\bar R_{xx}$ with $z$ between regions of low and strong shear than what is actually observed in the DNSs, where $\bar R_{xx}$ is roughly constant.

		\subsection{The stratified limit}
		
\begin{figure}[!ht]
			\begin{center}
				\includegraphics[width=0.8\textwidth]{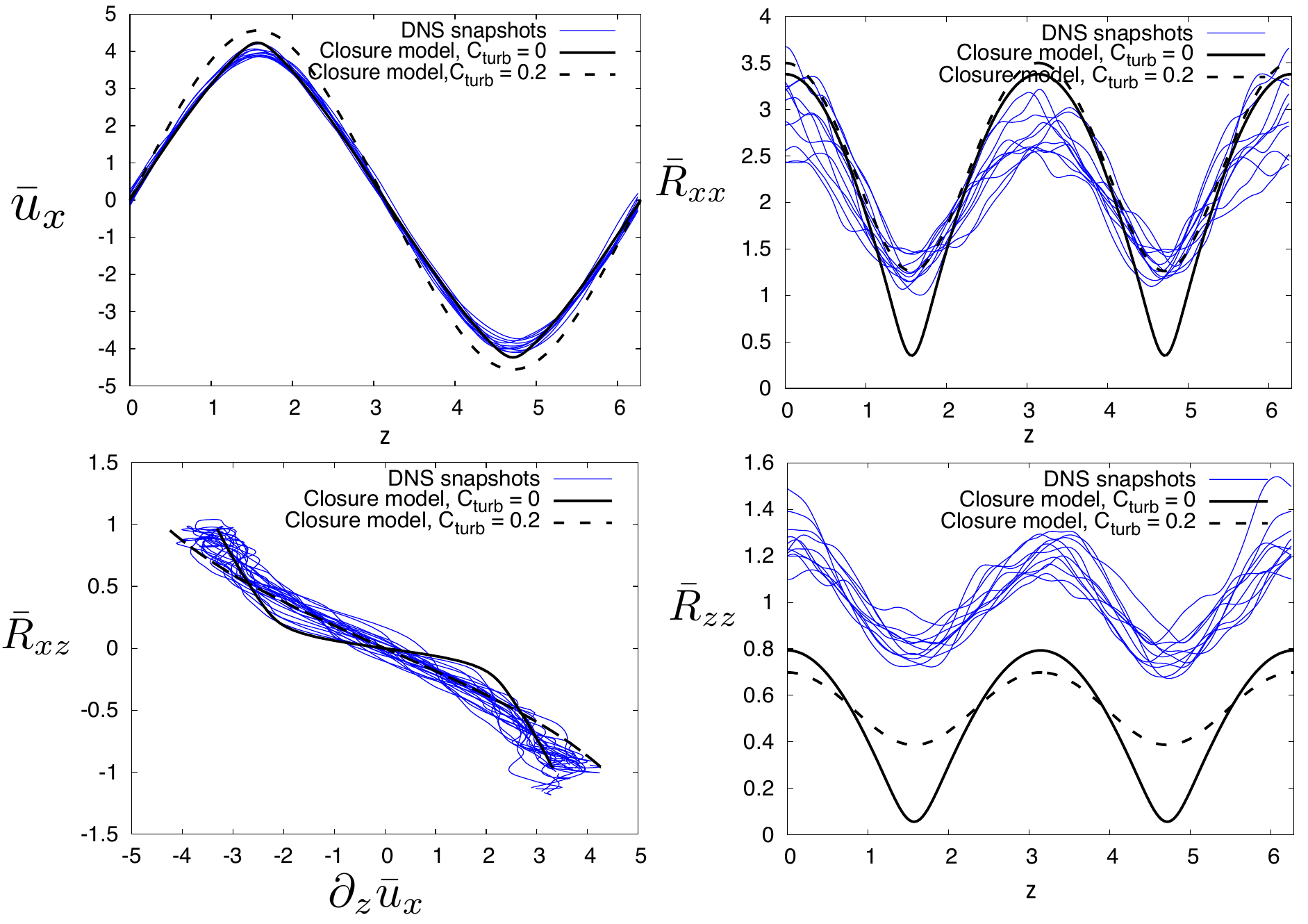}
				\caption{As in Fig. \ref{fig:closure10.01}, but for $\Ree_F = 100$, $\Ri_F \Pe_F = 1$ and $L = 0.9$.   }
				\label{fig:closure11}
			\end{center}
		\end{figure}

		Fig. \ref{fig:closure11} shows the same quantities as Fig. \ref{fig:closure10.01} for  a somewhat more strongly stratified case with $\Ree_F=100$ and $\Ri_F\Pe_F=1$. The value of $L$ for the closure model that best fits the data in this case is $L = 0.9$. The DNSs exhibit significantly different dynamics from the more weakly stratified case. The intrinsic variability of the system in time is now much smaller, which could have been expected since increasing the stratification stabilizes the perturbations. The mean flow remains in phase with the forcing, but is now somewhat more triangular, as pointed out by \citet{GaraudKulen16}. The energy in the perturbations (as measured by $\bar R_{xx}$ and $\bar R_{zz}$) drops significantly but again not completely in shear free regions. The closure model remains very good at predicting the shape of the mean flow $\bar u_x$,  and is also somewhat more successful at predicting both $\bar R_{xx}$ and $\bar R_{zz}$ than in the weakly stratified case. It still globally underestimates $\bar R_{zz}$, however, and still underestimates both $\bar R_{xx}$ and $\bar R_{zz}$  in the regions of no shear. Finally, the predicted variation of $\bar R_{xz}$ with the shearing rate $\partial_z \bar u_x$ is also much more nonlinear than in the DNSs, with the greatest discrepancy between the two taking place in regions of weak shear. 
			
		\subsection{The strongly stratified limit}

\begin{figure}[!ht]
			\begin{center}
				\includegraphics[width=0.8\textwidth]{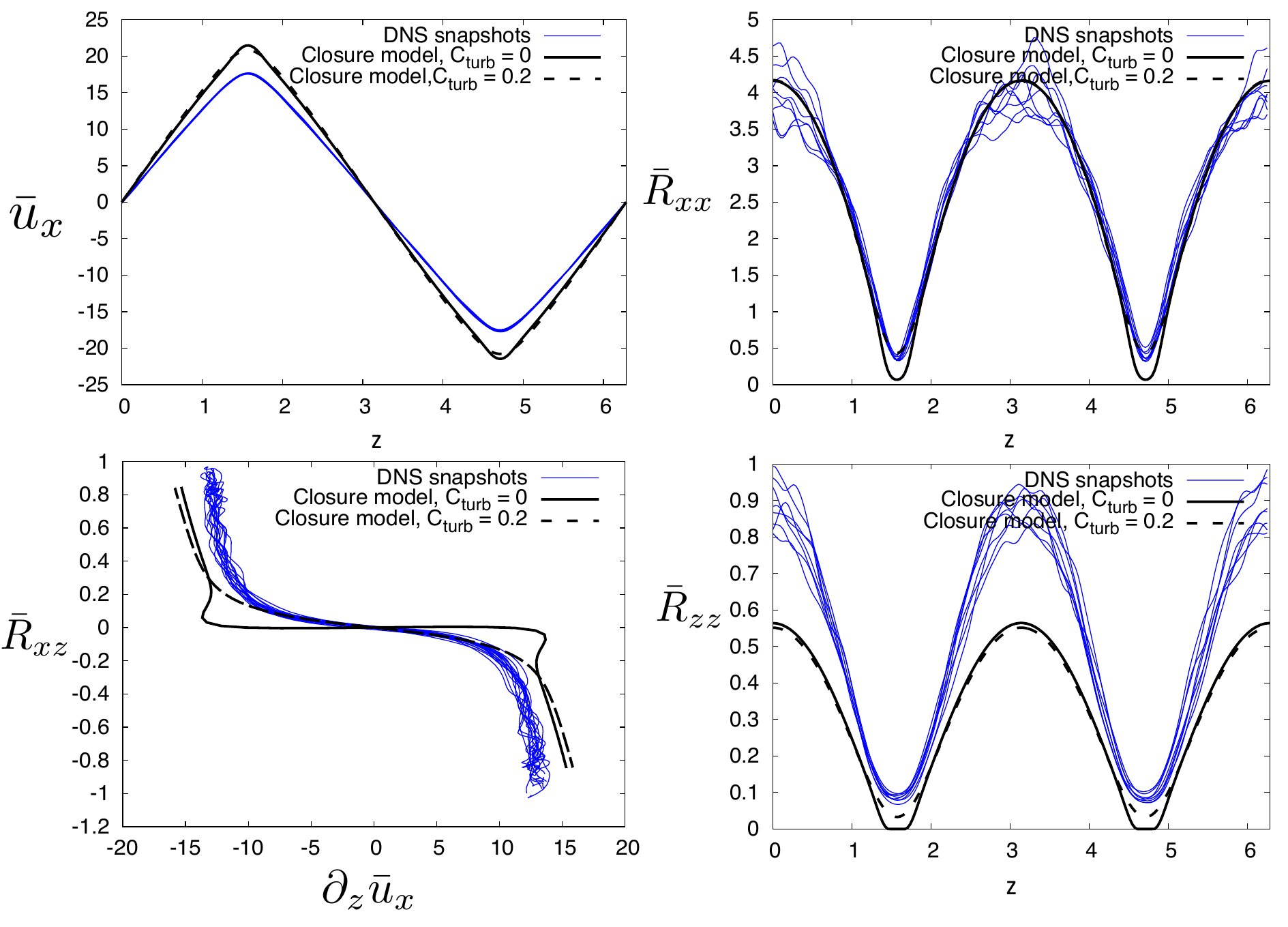}
				\caption{As in Fig. \ref{fig:closure1100}, but for $\Ree_F = 100$,  $\Ri_F \Pe_F = 100$ and $L = 0.34$.   }
				\label{fig:closure1100}
			\end{center}
		\end{figure}

Fig. \ref{fig:closure1100} shows the same quantities as Fig. \ref{fig:closure10.01} for a strongly stratified case with $\Ree_F=100$ and $\Ri_F\Pe_F=100$. The value of $L$ for the closure model that best fits the data in this case is $L = 0.34$. The intrinsic variability of the DNSs is now fairly minimal.  As discussed by \citet{GaraudKulen16}, the mean flow is very nearly triangular, with regions of nearly constant shear separated by very thin layers with no shear. The energy in the perturbations (as measured through $\bar R_{xx}$ and $\bar R_{zz}$) is close (but not exactly equal) to zero where the shear is null. 
The closure model is now overall quite good at predicting the DNS data. The predicted mean flow has the correct triangular shape, although its amplitude is slightly overestimated by the model. The model correctly captures the overall variation of both $\bar R_{xx}$ and $\bar R_{zz}$ with $z$, but continues to underestimates $\bar R_{zz}$ somewhat everywhere. The model also predicts a non-monotonous behavior for $\bar R_{xz} (\partial_z \bar u_x)$, which is not seen in the data.  

	\subsection{Adding a turbulent diffusivity}
	\label{sec:Cturb}

	Overall, we have found that this basic closure model performs adequately, though not as well as we had hoped for. It can reproduce the most important features of the data in the strongly stratified limit, but not so well in the weakly stratified limit. In particular, it has a tendency to underestimates the smoothness of the Reynolds stress profiles, as if it were missing a turbulent diffusion term that would act on stresses themselves. Indeed, the evolution equations for $\bar R_{ij}$ each contain a term in $\Ree^{-1} \partial^2_{zz} \bar R_{ij}$, which accounts for the microscopic diffusion of $\bar R_{ij}$. This term is essentially negligible for large enough Reynolds number, however, and therefore fails to account for the non-local effects we are trying to capture. For this reason, we propose to add a new term to the closure equations for $\bar R_{ij}$ that takes the form of a turbulent diffusivity, with a diffusion coefficient proportional to $L \bar R^{1/2}$ put forward purely on dimensional grounds. With this term, the Reynolds stress evolution equation becomes 
	\begin{eqnarray}
	\partial_t\bar R_{ij} + &\bar u_k\partial_k\bar R_{ij} +  \bar R_{ik}\partial_k\bar u_j+\bar R_{jk}\partial_k\bar u_i - \Ri\Pe (\bar f_i\delta_{jz} + \bar f_j\delta_{iz} ) - \left( \frac{1}{\Ree} + C_{\rm turb} \bar R^{1/2} L\right) \partial_{kk}\bar R_{ij} \nonumber \\&=-\frac{C_1\bar R^{1/2}}{L} \bar R_{ij} -\frac{C_2\bar R^{1/2}}{L}\left( \bar R_{ij}-\frac{1}{3}\bar {R}\delta_{ij}\right) -\frac{C_{\nu}\bar R_{ij}}{\Ree L^2}
	\label{eq:Cturbeq} 
	\end{eqnarray}
	where we have introduced a new model constant $C_{\rm turb}$, which is of order unity. Because the new term is only active for non-homogeneous shear flows, none of the results obtained regarding the local properties of the model discussed in Section \ref{sec:local} are affected.  
		
	We have found that, {\it without re-fitting $L$}, the value $C_{\rm turb} = 0.2$ provides a significantly better fit to the DNS results for the stratified ($\Ri_F \Pe_F = 1$) and strongly stratified ($\Ri_F \Pe_F = 100$) runs. The model continues to underestimate $\bar R_{zz}$, but now better captures the presence of mixing in the regions of no shear due to turbulent overshoot. For instance, the model predictions are much closer to the profile of $\bar R_{xz}(\partial_z \bar u_x)$ measured in the DNSs especially in the region of no shear. The same is true for $\bar R_{xx}(z)$ and $\bar R_{zz}(z)$. On the other hand, the agreement between the model and the data is now significantly worse for the run with $\Ri_F \Pe_F = 0.01$, where a much better fit would be obtained with a much smaller value of $C_{\rm turb} \simeq 0.01$. Of course, we could improve the agreement by re-fitting $L$ for these low $\Ri_F \Pe_F$ runs, but this would invalidate our argument that $L$ is proportional to the eddy scale $l_e$, and would defeat the purpose of having a physically-motivated model with reliable predictive power. Instead, we have simply demonstrated that the new model, with $C_{\rm turb} = 0.2$, performs fairly well for $\Ri_F \Pe_F \ge 1$, but should not be used for lower values of $\Ri_F \Pe_F$. 
		
	\section{Comparison of the time-dependent closure model predictions with DNSs of stratified periodic shear flows}
\label{sec:tanh}

	Having constrained the model parameters, and gained some experience using the proposed closure to reproduce the dynamics of very simple shear flows such as stratified plane Couette flows and Kolmogorov flows in a statistically-stationary state, we now investigate its ability to model more complex problems such as the simulations presented in Paper II. As in \citet{GaraudKulen16}, Paper II studies the nonlinear evolution of a body-forced diffusive shear flow by integrating equation (\ref{eq:momentumLPNpaper1}) in a triply-periodic domain. This time, the non-dimensional forcing function $\hat F(z)$ is chosen in such a way as to generate regions of strong shear adjacent to regions of weak shear, and would drive the following laminar flow in the absence of turbulent mixing:
	\begin{equation}
	U_L(z) =	\Ree_F \frac{\tanh\left[ a \sin\left( \frac{2\pi z}{L_z} \right) \right]}{\tanh(a)} \mbox{ where } L_z = \frac{2 \pi a  }{  \tanh(a)} ,
	\end{equation}
	and where $a$ is a shape parameter. Note how the domain size $L_z$ varies with $a$ in this model.  As discussed in Paper II, $U_L(z)$ tends to the Kolmogorov flow $\Ree_F \sin\left(z \right)$ with $L_z = 2\pi$ for $a \rightarrow 0$. As $a$ increases, the midpoint shear and the total amplitude of the flow remain the same, but increasingly large regions with little-to-no shear appear on either sides of the shear layer. As $a \rightarrow \infty$, $L_z \rightarrow \infty$, and $U_L(z)$ tends to the hyperbolic tangent profile $\Ree_F \tanh(z)$ instead (see Figure 1 of Paper II). The main finding of the DNSs of Paper II was that the weakly sheared regions undergo some form of mixing even if they locally do not satisfy Zahn's criterion $J {\rm Pr} < 0.007$. The extent to which they are mixed strongly depends on the stratification -- a result which prompted us to develop this closure model in the first place. 

        In what follows, we compare the closure model predictions against one particular DNS reported in Paper II, which has $\Ree_F = 100$, $\Ri_F \Pe_F = 10$ and $a = 2$, and which starts from initial conditions that are constructed from the laminar solution $U_L(z)$ plus small perturbations (referred to as Simulation A in Paper II). Rather than focussing on the ultimate statistically stationary state, we look at the full spatio-temporal evolution of the system.  Both the DNS and the closure model simulation are initialized to be very close to the equilibrium laminar flow $U_L(z)$ plus small perturbations of similar amplitudes (note that the perturbations are added to the full 3D velocity field in the DNSs, while they are added to the horizontally averaged Reynolds stress components $\bar R_{ij}$ in the closure model, so they cannot be exactly comparable). 
               
       We apply the closure model with the added turbulent diffusion term for the Reynolds stresses introduced in Section \ref{sec:Cturb}, and evolve the mean flow equation (\ref{eq:cm2}) with $\bar T = 0$, the Reynolds stress equations (\ref{eq:Cturbeq}), and the flux equations (\ref{eq:dcmt1}) and (\ref{eq:dcmt3}) in time from the initial conditions described above. We adopt the fiducial parameters, and take $L = 0.7$ constant throughout the domain. This particular value was not fitted, but instead was selected because this was the one determined to be the best fit for the corresponding Kolmogorov flow simulation with the same parameters (i.e. the same values of $\Ri_F \Pe_F  =10 $ and $\Ree_F = 100$) in the previous section, which has the same laminar mid-point shear and uses the same non-dimensionalization. Two cases are considered: $C_{\rm turb} = 0$ (ie. without turbulent diffusion of the Reynolds stresses), and $C_{\rm turb }= 0.2$ (with it). 
            
\begin{figure}[!ht]
			\begin{center}
				\includegraphics[width=\textwidth]{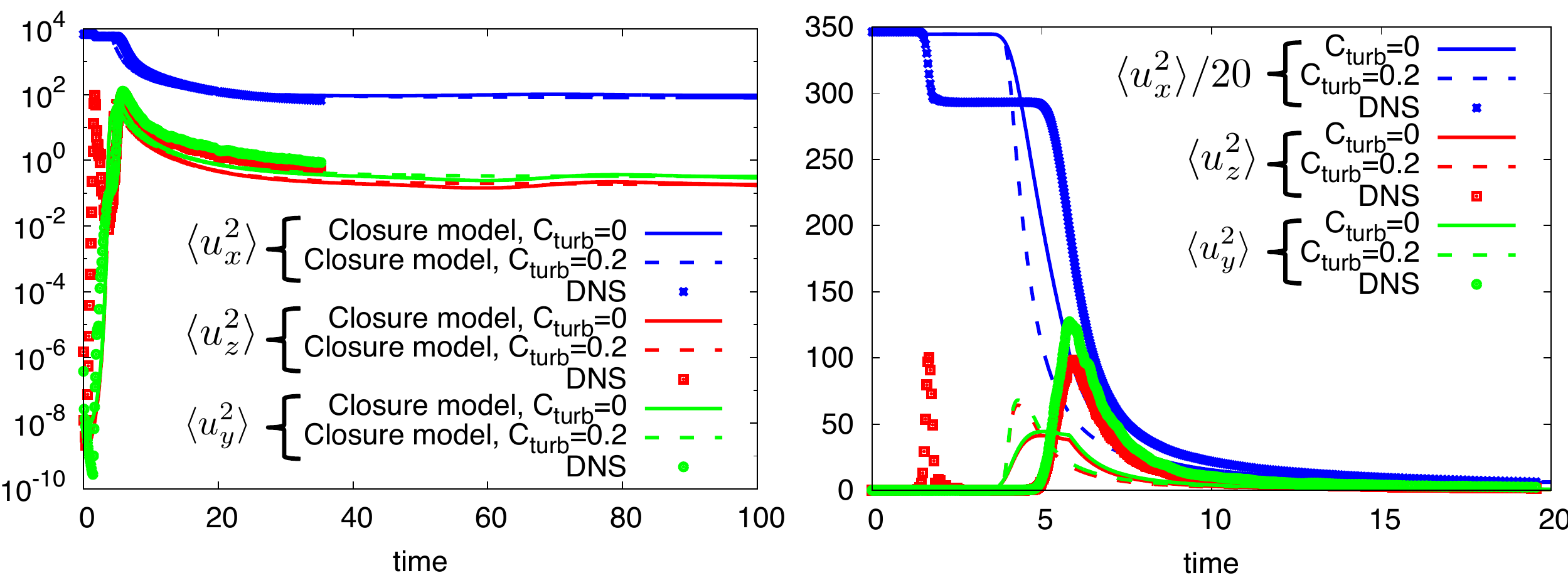}
				\caption{Comparison of the DNSs with the closure model. In both panels, the symbols correspond to the DNSs, the solid line to the closure model predictions without turbulent diffusion for the Reynolds stresses (i.e with $C_{\rm turb} = 0$) and the dashed line to the closure model predictions with turbulent diffusion (with $C_{\rm turb} = 0.2$). The left panel shows the evolution of $\langle u_x^2 \rangle$ (blue), $\langle u_y^2 \rangle$ (green) and $\langle u_z^2 \rangle$ (red) from $t = 0$ to $t = 100$, while the right panel shows corresponding quantities are early times. }
				\label{fig:CompareEnergies}
			\end{center}
		\end{figure}

        Figure \ref{fig:CompareEnergies} shows a comparison of volume averaged $\langle u_x^2 \rangle$, $\langle u_y^2 \rangle$, $\langle u_z^2 \rangle$ observed in the DNSs, with the corresponding quantities extracted from the closure model, namely 
       \begin{equation}
       \langle u_x^2 \rangle = \frac{1}{L_z} \int_0^{L_z} (\bar u_x^2 + \bar R_{xx}) dz \mbox{   ,   } \langle u_y^2 \rangle = \frac{1}{L_z} \int_0^{L_z} \bar R_{yy} dz \mbox{   ,   } \langle u_z^2 \rangle = \frac{1}{L_z} \int_0^{L_z} \bar R_{zz} dz.
	\end{equation}

We see that the $C_{\rm turb} = 0$ and $C_{\rm turb} = 0.2 $ cases are qualitatively similar to one another, and only differ in the details (see below). Both correctly capture many (but not all) of the observed trends from the DNSs.  For instance they are fairly successful in capturing the overall evolution of the kinetic energy of both streamwise and spanwise fluid motions, at least qualitatively. However, they fail to account for a first burst of activity in the vertical motions, associated with a first decrease in the energy of streamwise motions. This feature is seen in the 3D DNSs, but not in either of the closure model runs. Paper II attributes this first burst in the DNSs to a purely 2D instability (with motion in the $x-z$ plane that is invariant in the spanwise direction), which only later becomes unstable to 3D perturbations -- this is a well known feature of the transition to turbulence in stratified shear flows \citep{PeltierCaulfield03}. Since the closure model on the other hand intrinsically disallows purely 2D motion (because of the assumption of isotropization), it only properly models the 3D phase of the instability, but misses the initial 2D phase. Even so, the time at which the instability becomes fully developed, and its saturation amplitude, are relatively well captured by the model.

Looking at this comparison from a more quantitative point of view, we see that the closure model with $C_{\rm turb} = 0.2$ performs somewhat better than the case with $C_{\rm turb} = 0$ on longer timescales. In particular the $C_{\rm turb} = 0$ case fails to reach a steady state, and undergoes instead a series of quasi-periodic oscillations which are not present in the DNSs. Inspection of the results show that these oscillations are due to the $C_{\rm turb} = 0$ model being unable to pin down the edges of the laminar region, which periodically move towards and away from the midpoint of the shear layer with time. This does not happen in the  $C_{\rm turb} = 0.2$ case, presumably because the regions of low shear are never fully quiescent. 

From here on we only consider the case with turbulent diffusion for the Reynolds stresses, with $C_{\rm turb} = 0.2$, since its behavior is more regular. To see how well that model captures the vertical profiles of the mean flow and of the various Reynolds stress tensor components, we compare them to one another at times $t = 10$ (a little after the main turbulent mixing event has taken place), $t = 20$ (when the system is starting to relax to a statistically stationary state), and $t = 30$ (where it is close to being in a statistically stationary state). We see that the closure model captures the general properties of the solutions relatively well (i.e. within $\sim 30\%$) except as usual for $\bar R_{zz}$ (and $\bar R_{yy}$, not shown) which is underestimated by a factor of about 2.

\begin{figure}[!ht]
			\begin{center}
				\includegraphics[width=0.8\textwidth]{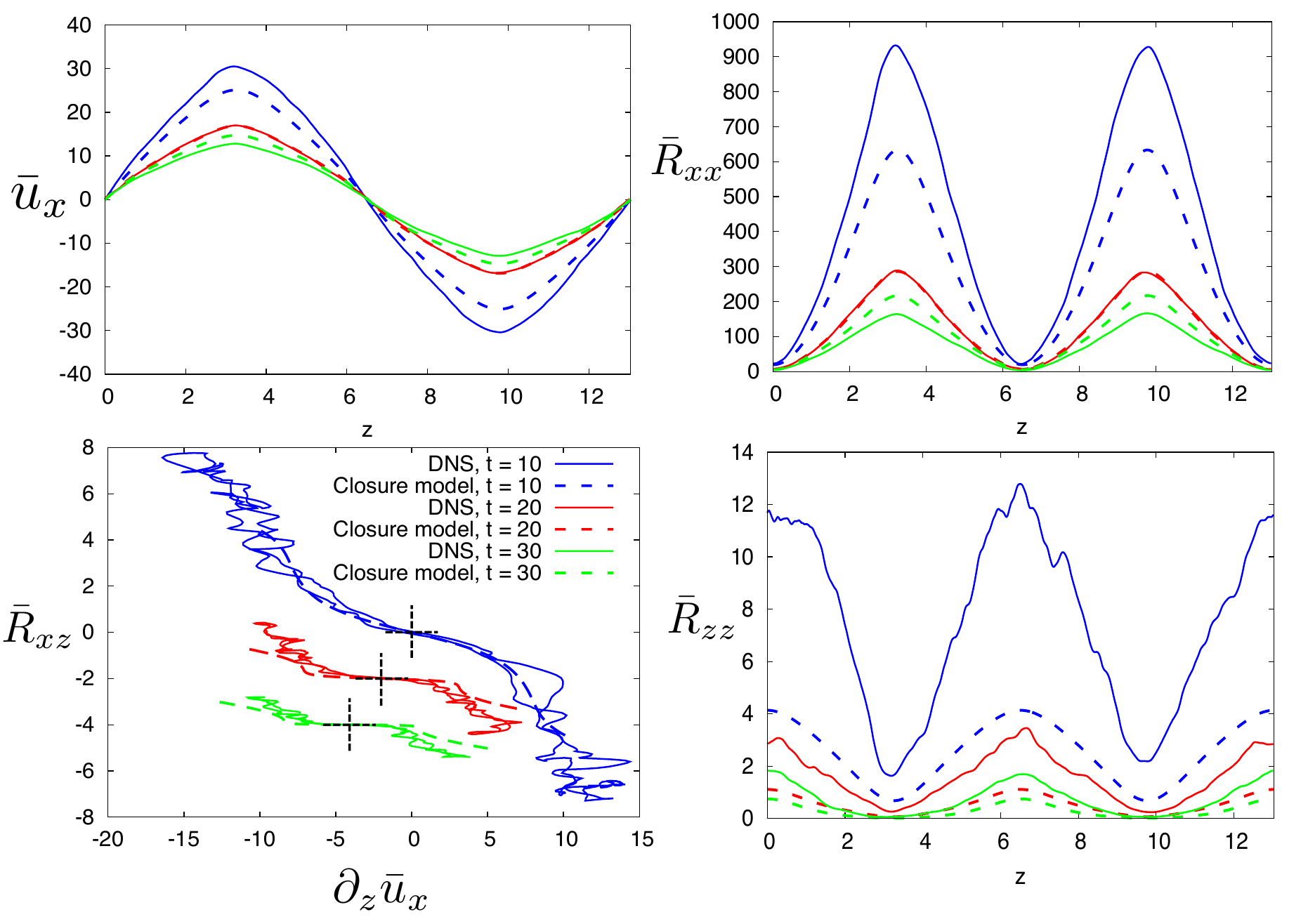}
				\caption{Comparison of the DNSs with the closure model for $C_{\rm turb} = 0.2$ and $L = 0.7$, at times $t = 10$, $t = 20$ and $t = 30$. In each panel, the solid line corresponds to the DNSs, and the dashed line corresponds to the model. In the bottom left panel, $\bar R_{xz}$ is shown as a function of $\partial_z \bar u_x$, although the axes only correspond to the data for $t = 10$. The data for $t  = 20$ and $t  =40$ are shifted down and to the left by $2$ and $4$ units respectively, to avoid overlap of the lines. The cross marks the shifted origin. The model captures the spatio-temporal evolution of the data within 20\% - 30\%, except for $\bar R_{zz}$ (and $\bar R_{yy}$) where it underestimates it by a factor of about 2. }
				\label{fig:Comparetanh}
			\end{center}
		\end{figure}

To summarize, we find that the closure model thus far performs adequately, reproducing the spatio-temporal evolution of both $\bar u_x$ and of $\bar R_{ij}$ typically within 20-30\% accuracy, except for $\bar R_{yy}$ and $\bar R_{zz}$ which it systematically underestimates by a factor of about 2. 

       \section{How to select $L$} 
       \label{sec:L}

      Up until this point, the closure model has relied on the user to manually input the eddy scale $L$, which is not practical for astrophysical applications where one would rather automate the process. Furthermore, this scale was assumed to be constant in the domain and in time, which may not necessarily be the case if the shear varies significantly. To address both issues we now propose a simple prognostic equation for $L$. Note that in this section we use the same non-dimensionalization as in Sections \ref{sec:Kolmo} and \ref{sec:tanh}, but the results are easy to generalize to other non-dimensionalizations.

Fitting the data from Figure \ref{fig:Lcompare} \citep[which shows the values of $L$ that best fit the sinusoidally-forced simulations of][]{GaraudKulen16}, we find that $L$ is well-approximated by 
       \begin{equation}
       L_{\rm theor} = \frac{1}{l_{\rm max}^{-1} + L_{\rm Z}^{-1}}  \mbox{    where   }  L_{\rm Z} = (J \Pe_S)^{-1/2} ,
       \label{eq:Ltheor}
       \end{equation}
       where $l_{\rm max}$ is the size of the shear layer (which was $\pi$ for the simulations shown in Figure \ref{fig:Lcompare}), and $J {\rm Pe}_S = {\rm Ri}_F {\rm Pe}_F / |S|$ is the local Richardson-P\'eclet number based on the local shearing rate $S$ (written in the non-dimensionalization appropriate for Figure \ref{fig:Lcompare}).  This formula guarantees that $L_{\rm theor} \rightarrow l_{\rm max}$ when stratification is negligible, and that $L \rightarrow  L_{\rm Z}$ when stratification is dominant. The quantity $L_{\rm Z}$ should by definition be equal to 1.5 times the Zahn scale, which is equal to $l_{\rm Z} = \sqrt{ \frac{ (J {\rm Pe})_c }{ J \Pe_S} }$ when written nondimensionally so 
       \begin{equation}
       L_{\rm Z}  = 1.5 \sqrt{ \frac{ (J {\rm Pe})_c }{ J \Pe_S} },
       \end{equation}
       where $(J {\rm Pe})_c$ is a constant of order unity. 
     Fitting this expression to the data from Figure \ref{fig:Lcompare}, we find that the result coincidentally happens to take the very simple form $ L_{\rm Z} \simeq (J \Pe_S)^{-1/2}$, which implies that $(J {\rm Pe})_c  \simeq 0.44$ (see also Paper II). Since the non-dimensionalization is the same for the DNSs of \citet{GaraudKulen16} (from which Figure \ref{fig:Lcompare} was created) and the simulations of Paper II, we can expect that $L_{\rm Z}$ and therefore $L_{\rm theor}$ will be given by equation (\ref{eq:Ltheor}). 
       
  	Equation (\ref{eq:Ltheor}) can be used in a number of different ways. For instance, if a typical value of the actual shearing rate of the turbulent flow is known a priori, and if the system is close to being in a statistically stationary state, then one can simply set $L$ to be constant and equal to the corresponding $L_{\rm theor}$ at every point in space and for all times. This is essentially what we have been doing until this point (even though the system was not always close to being statistically stationary). A more sophisticated alternative would be to let $L_{\rm theor}$ evolve in response to the spatiotemporal variation of the shearing rate. One can also  let $l_{\rm max}$ vary with position in this expression,  which would be the case for instance if the system was wall-bounded (in which case $l_{\rm max}$ would be the distance to the nearest wall). It is worth noting however that the scale $L_{\rm theor}$ thus created is very sensitive to the value of the local shear $S$,  which poses two problems. First, this sensitivity tends to be a destabilizing factor in any numerical scheme selected for the  evolution of the closure model, and second, using $L = L_{\rm theor}$ is somewhat unrealistic since in practice one would not expect $L$ to vary on a lengthscale significantly shorter than itself. 
	
	We therefore propose as a third alternative the following evolution equation for $L$: 
       \begin{equation}
       \partial_t L  = -  \frac{\bar R^{1/2}}{L} (L - L_{\rm theor}) + \left( \frac{1}{\Ree} + C_{\rm turb} \bar R^{1/2} L \right) \partial_{zz} L
       \label{eq:Lmodel}
       \end{equation}
       where $\Ree = \Ree_F$ when applying the model in the non-dimensionalization specific to this section. 
      The first term drives $L$ toward $L_{\rm theor}$ on the eddy turnover time $L / \bar R^{1/2}$, while the second term is a turbulent diffusion term (where the $1/\Ree$ is there merely to regularize the diffusion coefficient in the limit where $\bar R^{1/2} = 0$) which smoothes out the spatial variability of $L$. Note that we could have introduced additional model constants in front of the first term, and we could have selected a different constant instead of keeping $C_{\rm turb}$ in the second term. However, we have checked that these would not substantially change the outcome, so we prefer to leave the model as simple as possible. 

\begin{figure}[!ht]
			\begin{center}
				\includegraphics[width=\textwidth]{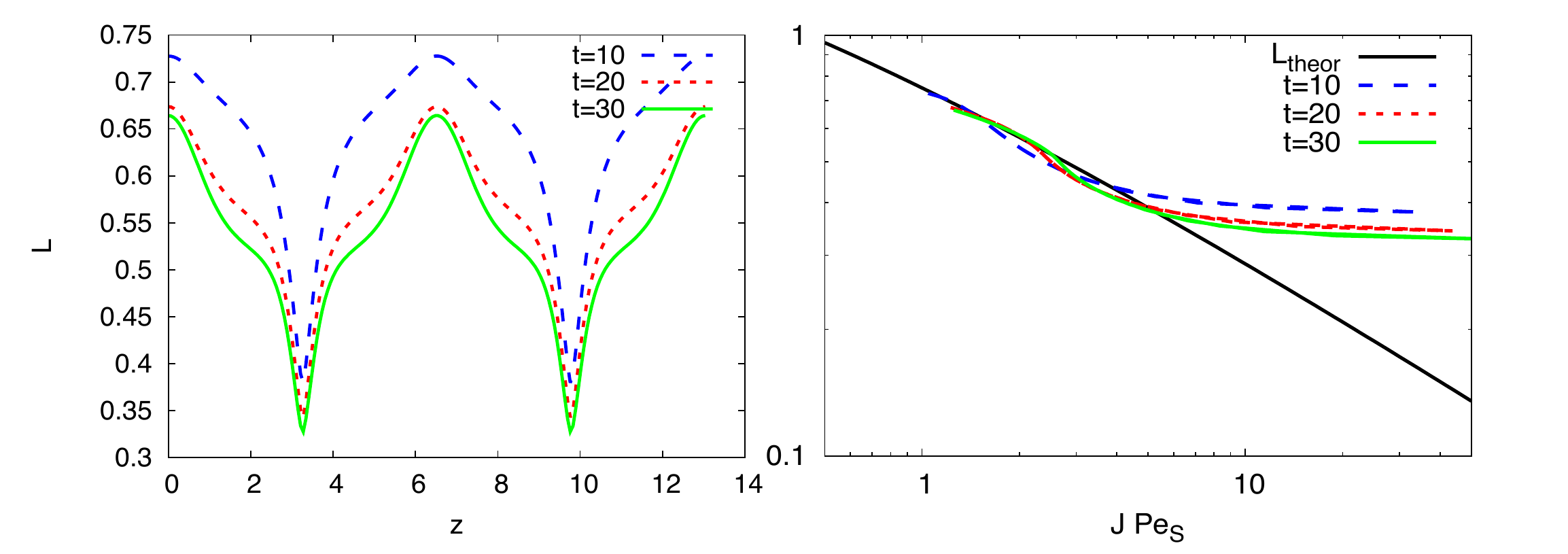}
				\caption{Evolution of $L$ as a function of time according to the new model given by equation (\ref{eq:Lmodel}). The left panel shows $L(z)$ at three selected times. The right panel shows $L$ as a function of $J \Pe_S$, at the same three times. $L$ agrees overall with $L_{\rm theor}$ (see equation \ref{eq:Ltheor}), but deviates from it in regions of very low shear, where it is controlled by turbulent diffusion from nearby regions. }
				\label{fig:Lmodel}
			\end{center}
		\end{figure}

      The complete proposed closure equations are therefore the mean flow equation (\ref{eq:cm2}), the Reynolds stress equations (\ref{eq:Cturbeq}), the flux equations (\ref{eq:dcmt1}) and (\ref{eq:dcmt3}) together with (\ref{eq:Lmodel}) for $L$. We evolve them in time with the same initial conditions as in the constant $L$ case, taking in addition $L(z,0) = L_{\rm theor}(z)$ at $t = 0$. The results are nearly identical to those presented in Figures \ref{fig:CompareEnergies} and \ref{fig:Comparetanh}, and are not worth showing: the model performs identically well (or poorly, depending on one's perspective) in reproducing the data. As such, equation (\ref{eq:Lmodel}) is not meant to improve the accuracy of the model, but rather, to give a simple, stable, and automatic way of predicting $L$. Figure \ref{fig:Lmodel} shows $L$ at times $t = 10$, $t = 20$ and $t = 30$, as a function of $z$ (left), and as a function of $J {\rm Pe}_S$ (right). This second plot shows that $L$ thus constructed is roughly equal to $L_{\rm theor}$ everywhere in the domain, except in regions of very low shear (high $J{\rm Pe}_S$) where it is nearly constant. This can be understood by noting that when $S \rightarrow 0$, the evolution of $L$ is dominated by turbulent diffusion from nearby regions. 

It is worth noting that in this particular simulation where the stratification is fairly large, the selected value of $l_{\rm max}$ (which describes the intrinsic lengthscale associated with the shear) is irrelevant. For more weakly stratified cases, the user must specify $l_{\rm max}$, and could select it either by considering a shear or forcing scaleheight (for body-forced cases), or the domain size and/or distance to the nearest boundary (for wall-bounded cases). However, since we have also shown that the model does not perform particularly well for low stratification anyway, the uncertainty in choosing $l_{\rm max}$ in this limit is perhaps not a particularly pressing problem.

\section{Conclusion}
\label{sec:ccl}

\subsection{Summary} 

In this series of papers (Paper I, Paper II and this work), as well as in previous work \citep{Garaudal15,GaraudKulen16} we have provided an in-depth investigation of turbulent mixing (of both chemical species and momentum) driven by the diffusive shear instability, excluding the effects of chemical stratification, rotation and magnetic fields. The diffusive shear instability only exists in the low Prandtl number environment inherent to stars and gas giant planets, since it requires thermal diffusion to be strong (low P\'eclet number) while viscous diffusion is weak (high Reynolds number). Using DNSs, we were able to test the nonlinear instability criterion (\ref{eq:Zahncrit}) and local mixing model (\ref{eq:ZahnDturb}) proposed by \citet{Zahn1974,Zahn92}. We found both to be correct in the context of numerical experiments specifically designed to reproduce the conditions under which Zahn's model should apply, namely (i) that the P\'eclet number be small \citep{GaraudKulen16}, and (ii) that the turbulence properties are inherently local with a turbulent eddy scale that is much smaller than the system scale or the shear lengthscale \citep[see Paper I and the works of][]{PratLignieres13,PratLignieres14,Pratal2016}. On the other hand we also showed that the model fails when this locality condition fails (Paper I, Paper II), which happens when the eddy scale becomes large compared with the shear lengthscale or the system scale. This can be problematic in two cases: (a) when the stratification is weak, in which case the intrinsic turbulent eddy scale (which is theoretically given by the Zahn scale) becomes very large, or (b) in modeling mixing near and beyond the edge of the theoretically unstable region, when the eddy scale is commensurate with the distance to that edge. In Paper II, we also showed that the nonlinear nature of the instability in the more strongly stratified case must be treated with care, as it gives rise to hysteresis -- a possibility that cannot be accounted for in Zahn's model. 

In this paper we have therefore put forward a more realistic model for mixing by diffusive shear instabilities that can (at least in principle) self-consistently address all of these problems. This model is based on the second-order turbulence closure paradigm first proposed by \citet{Ogilvie2003}, and later applied to unstratified shear flows by \citet{GaraudOgilvie2005}. It is expressed non-dimensionally, using a unit velocity $U_0$, a unit length $L_0$, and depends on three non-dimensional parameters, namely the system-scale Richardson, Reynolds and P\'eclet numbers ${\rm Ri}$, ${\rm Re}$ and ${\rm Pe}$ given in equations (\ref{eq:Ri})-(\ref{eq:Pe}). 
We have derived the closure equations, and applied the asymptotic limit of low P\'eclet number to obtain a first set of model equations given in (\ref{eq:dcm1})-(\ref{eq:dcmt3}). By comparing the model predictions with DNSs from Paper II, we noted the need for a turbulent diffusivity term in the Reynolds stress equations, and added it. When applied to plane parallel shear flows (with the mean flow in the horizontal $x$ direction and shear in the vertical $z$ direction) one can extract a reduced set of coupled equations :  
\begin{eqnarray}
                          &&\begin{split}
                          &&\partial_t \bar u_x + \partial_z \bar R_{xz}= \frac{1}{\Ree}\partial^2_{zz} \bar u_x \left(  + \hat F(z) \right) \nonumber  ,
                         \end{split}
                         \\
			&&\begin{split}
			\partial_t\bar R   +  2\bar R_{xz}\partial_z\bar {u}_x  -  2\Ri\Pe\bar f_z    = \left( \frac{1}{\Ree } + C_{\rm turb} \bar R^{1/2} L \right) \partial^2_{zz} \bar R  -\frac{C_1 \bar R^{3/2}}{L}  -\frac{C_{\nu}\bar R}{\Ree L^2}   \nonumber ,
			\end{split}	
			\\
			&&\begin{split}
			\partial_t\bar R_{zz} -  2\Ri\Pe\bar f_z   =  \left( \frac{1}{\Ree } + C_{\rm turb} \bar R^{1/2} L \right) \partial^2_{zz}\bar R_{zz} -(C_1+C_2)\frac{\bar {R}^{1/2}}{L} \bar R_{zz}  + \frac{C_2\bar {R}^{3/2}}{3L} -\frac{C_{\nu}\bar R_{zz}}{\Ree L^2}  \nonumber,
			\end{split}	
			\\
			&&\begin{split}
			\partial_t\bar R_{xz} +  \bar R_{zz}\partial_z\bar {u}_x- \Ri \Pe \bar f_x = \left( \frac{1}{\Ree } + C_{\rm turb} \bar R^{1/2} L \right) \partial^2_{zz}\bar R_{xz} -(C_1+C_2)\frac{\bar {R}^{1/2}}{L} \bar R_{xz}  -\frac{C_{\nu}\bar R_{xz}}{\Ree L^2}  \nonumber,
			\end{split}	
			\\
			&&\begin{split}
			\bar R_{xz} - \frac{1}{2}\partial^2_{zz}\bar f_x=-\frac{1}{2L^2}C_{\nu \kappa}\bar f_x  \nonumber,
			\end{split}
			\\
			&&\begin{split}
			\bar R_{zz} - \frac{1}{2}\partial^2_{zz}\bar f_z=-\frac{1}{2L^2}C_{\nu \kappa}\bar f_z \label{eq:ccldcmt3}.
			\end{split}
			\end{eqnarray}
for the evolution of the mean flow $\bar u_x$ (where the overbar is a horizontal average), of the components of the Reynolds stress tensor $\bar R_{zz}$ and $\bar R_{xz}$, for (twice) the turbulent kinetic energy $\bar R = \bar R_{xx} + \bar R_{yy} + \bar R_{zz}$, and for the temperature fluxes $f_x = \bar F_{x} / {\rm Pe}$ and $f_z = \bar F_z / {\rm Pe}$. Other components of the stress tensor can be obtained if desired, but do not have to be computed if not.  The unstratified limit ($\Ri \Pe = 0$) with $C_{\rm turb} = 0$ recovers the model of \citet{GaraudOgilvie2005}. 

The model can easily be exported to a spherical coordinate system if desired, to study shear-induced mixing in stars. However, we caution the reader against using it to model rotational shear indiscriminately, since the effect of rotation on the dynamics of diffusive shear instabilities has not been studied yet (this will be the subject of a future paper). To be precise, we believe that the model should almost certainly apply as is if the Rossby number based on the eddy scale $L$ is very large (i.e. if ${\rm Ro} = \bar R^{1 /2}/ 2 \Omega L \gg 1$, where $\Omega$ is the mean non-dimensional rotation rate of the region considered), but not if ${\rm Ro}$ becomes of order unity or smaller. 

The set of equations (\ref{eq:ccldcmt3}) contains 5 closure parameters $C_1$, $C_2$, $C_\nu$, $C_{\nu\kappa}$ and $C_{\rm turb}$ and relies on the user to propose a model for the turbulent eddy scale $L$. In their local form (i.e. ignoring all spatial derivatives except $\partial_z \bar u_x$ which is assumed constant), and assuming steady state, the closure model equations are algebraic and can be solved analytically or semi-analytically. The properties of this local model have been analyzed, and recover some of the known properties of diffusive shear instabilities, including the fact that $J {\rm Pe}_L$ should be smaller than a constant of order unity for instabilities to proceed (where $J$ is the gradient Richardson number and ${\rm Pe}_L$ is the eddy-scale P\'eclet number) \citep{Townsend58,Dudis1974,Zahn1974}. It also emphasizes the crucial role of the parameter $J {\rm Pr}$ (where Pr is the Prandtl number) in controlling the dynamics of the system, as in Zahn's model \citep{Zahn1974}. 

Comparison of the local closure model predictions with DNSs from Paper I  shows very good agreement, and can be used to fit the first four model constants. We therefore propose that 
\begin{equation}
C_1 \simeq 0.41,  \quad  C_2 \simeq 0.54,   \quad C_{\nu\kappa} \simeq 10   \mbox{   and   } C_\nu \simeq 15, 
\label{eq:cclparams}
\end{equation}
with errors of order 20\% for $C_1$, $C_2$ and $C_\nu$, and of order $50\%$ for $C_{\nu\kappa}$. These errors were estimated not only from comparison with the data from Paper I, but also by comparison with the constant values obtained from fitting the closure model against entirely different kinds of numerical and laboratory experiments, e.g. pipe flow and Taylor-Couette flows \citep{GaraudOgilvie2005} and Rayleigh-B\'enard convection \citep{Garaudal2010}.

By comparing the full closure model predictions with DNSs from \citet{GaraudKulen16}, we also found that the best fit is obtained when the eddy scale $L$ is proportional to the actual turbulent eddy scale $l_e$, measured using the vertical autocorrelation of the spanwise flow. In Paper I, we discovered that $l_e$ is directly related to the Zahn scale in the strongly stratified limit, and to the system scale in the weakly stratified limit. This then suggests that a simple theoretical model for $L$ can generally be constructed as follows: 
\begin{equation}
       L_{\rm theor} = \frac{1}{ l_{\rm max}^{-1} + (1.5l_{\rm Z})^{-1}}  , 
       \label{eq:cclLtheor}
       \end{equation}
where $l_{\rm max}$ is the outer scale (i.e. the overall size of the shear layer, or the shear lengthscale, or the distance to the wall in wall-bounded shear flows), and $l_{\rm Z} = ((J{\rm Pe})_c |S| / \Ri\Pe)^{1/2}$ is the Zahn scale (where $S = d\bar u_x/dz$ is the local shearing rate and $(J\Pe)_c$ is a constant of order unity), both expressed in units of $L_0$. The constant $(J\Pe)_c$ is of order unity, and can be fitted to the data if data is available as we have done in Section \ref{sec:L} for one particular set of DNSs, or taken to be equal to one otherwise. 
 
As proposed, the set of equations (\ref{eq:ccldcmt3}) together with (\ref{eq:cclparams}) and (\ref{eq:cclLtheor}), with $C_{\rm turb} = 0$, can adequately model the dynamics of diffusive shear flows under a wide range of parameters (with the caveats discussed below) and for different kinds of setups. However, the model is not particularly stable numerically, and is prone to unphysical small-amplitude oscillations associated with its inability to pin down the edges of the turbulent region adequately. Setting $C_{\rm turb}$ to a small value, e.g. $C_{\rm turb} = 0.2$, solves that problem without affecting the other results much as long as $\Ri \Pe$ is of order unity or larger (for smaller $\Ri\Pe$, a model with $C_{\rm turb} \equiv 0$ fares better). Similarly, letting $L$ evolve with time according to (\ref{eq:Lmodel}) rather than merely setting $L = L_{\rm theor}$ also improves the stability of the numerical scheme, without noticeably changing the ultimate results. This is important for stellar evolution calculations, where stability is essential.  

Finally, note that while we have so far ignored the possibility of chemical stratification and chemical mixing, this closure model can in fact easily be used to compute a turbulent diffusivity $D_{\rm turb}$ for chemical species (assuming they are a passive tracer, i.e. assuming they do not contribute to the buoyancy much). Indeed, we have shown in Paper I \citep[see also][]{Pratal2016} that $D_{\rm turb} \simeq \nu_{\rm turb}$ where $\nu_{\rm turb}$ is the turbulent viscosity. Since we can compute $\nu_{\rm turb}$ from our model as
\begin{equation}
\nu_{\rm turb} = - \frac{\partial \bar R_{xz}}{\partial S} ,
\end{equation}
$D_{\rm turb}$ can similarly be obtained. 
   
\subsection{Model caveats and future work} 

The proposed model, as demonstrated in Sections \ref{sec:local} through \ref{sec:L}, can be used to obtain qualitatively reliable predictions for the complex spatio-temporal behavior of diffusive stratified shear flows across a vast region of parameter space. For instance, it correctly captures the multiple evolutionary timescales associated with the flow, can correctly identify the edge of a mixed layer, as well as predic mixing beyond that edge. It also naturally captures the existence of multiple statistically stationary states, which can lead to hysteresis depending on initial conditions, a property of these shear flows that was first discussed in Paper II. It {\it qualitatively} captures the profile of the mean flow and of each component of the Reynolds stress tensor, but with varying degrees of {\it quantitative} accuracy.

In particular, the model is not particularly good in quantitatively reproducing the system dynamics in the limit of very low stratification. It does not correctly capture the partitioning of energy between the various components of the stress tensor (e.g. under-predicting $\bar R_{zz}$  and overpredicting $\bar R_{xx}$), and unless $C_{\rm turb} = 0$, does not accurately model the mean flow and momentum transport either. This may not be too much of a problem, as the very low stratification limit is likely not particularly prevalent in stars, but this limitation should be born in mind by the user.
When stratification increases, however, the model fares significantly better, and is able to predict the mean flow within $\sim 20\%$. It much more accurately predicts $\bar R_{xx}$ as well. Curiously, it continues to systematically underestimate the amplitude of $\bar R_{zz}$ by a factor of about 2, even though the shape is accurately predicted. Its predictive power for the shape of $\bar R_{xz}(S)$ (whose derivative is the turbulent viscosity) is relatively good as well. 

The discrepancies between the model predictions and the DNSs must necessarily be due to some incomplete modeling of the triple correlations, pressure-strain correlations, and large-scale diffusive terms by the proposed closure. We have looked into further additions to the model, to attempt to solve the problem. In particular, given that the pressure satisfies the (dimensional) Poisson equation
\begin{equation}
\frac{1}{\rho_0} \nabla^2 p = - \nabla \cdot (\bu \cdot \nabla \bu) + \alpha g \frac{ \partial T }{\partial z},
\end{equation}
we see that the pressure-strain correlation terms $\overline {u_j \partial_i p }$ in (\ref{eq:cm4}) will contain further triple-correlation terms in the unstratified limit (where $T = 0$), but may also contain correlations between $u_j$ and $T$ in the stratified case, which had so far been ignored. We are presently looking into the possibility of adding new closure terms proportional to the heat flux in the Reynolds stress equation. Preliminary results are encouraging and will be discussed in a future publication.

In general, however, our primary motivation is to create a closure that follows the same guiding principles as those outlined in the original work of \citet{Ogilvie2003}: 
sufficiently simple to be used in stellar evolution calculations, only containing closure terms with a simple physical explanation, and calibrated against DNSs. With this,  
we have shown its ability to predict the vertical profiles of the mean flow and of the stress tensor within a factor of order unity in the worse case scenario, but often much more precisely. While not perfect, this is still much better than the vast majority of mixing prescriptions used in stellar astrophysics, which have rarely or never been tested numerically. 

Finally, and as discussed in Paper II, there are significant limitations to the applicability of this model that the user should bear in mind. First, it does not take into account the effect of composition in the buoyancy of the material, and should therefore not be applied in regions where the density gradient due to composition is of the order of, or exceeds, the density gradient due to temperature stratification. Secondly, the validity of this model has only been demonstrated in the limit of low P\'eclet number. Whether any of it still holds for larger P\'ectlet number flows remains to be determined. Third, the model is not valid in regions where the Rossby number Ro = $|S|/\Omega$ is small, i.e. in the limit of rapid rotation. Since the shear $S$ is often dominated by rotational shear in stars, one should be particularly careful in checking whether this condition is met before applying the model. We are presently working to expand our understanding of stratified shear instabilities in all three of these directions. 

\acknowledgments
L.K. and P.G. gratefully acknowledge funding by NSF AST-1517927. The simulations were run on the Hyades supercomputer at UCSC, purchased with an NSF MRI grant. The PADDI code used was generously provided by S. Stellmach. We thank D. Gagnier and J. Verhoeven for access to the result of their DNSs. We also warmly thank the IDEX initiative at Universit\'e F́\'ed́\'erale Toulouse Midi-Pyr\'en\'ees (UFTMiP) for making this collaboration possible.

%
%\clearpage
%\bibliography{NSF-bib}

\providecommand{\noopsort}[1]{}\providecommand{\singleletter}[1]{#1}%

\end{document}